\documentclass{emulateapj}

\newcommand{\kms}{\,km\,s$^{-1}$} \newcommand{\sqcm}{\,cm$^{-2}$}  
\newcommand{\fuse}{\emph{FUSE}}  \newcommand{\hst}{\emph{HST}}
\newcommand{\os}{\ion{O}{6}}     \newcommand{\ct}{\ion{C}{3}}
\newcommand{\hi}{\ion{H}{1}}     \newcommand{\nf}{\ion{N}{5}}
\newcommand{\cf}{\ion{C}{4}}     \newcommand{\sif}{\ion{Si}{4}}
\newcommand{\sit}{\ion{Si}{3}}   \newcommand{\siw}{\ion{Si}{2}}
\newcommand{\cw}{\ion{C}{2}}     \newcommand{\ha}{H$\alpha$}
\newcommand{\oi}{\ion{O}{1}}     \newcommand{\few}{\ion{Fe}{2}}
\newcommand{\he}{HE~0226--4110}  \newcommand{\pg}{PG~0953+414}    
\newcommand{\tm}{\tablenotemark} \newcommand{\tn}{\tablenotetext}
\newcommand{\hvo}{high-velocity \ion{O}{6}}
\newcommand{\hva}{high-velocity absorption}
\newcommand{\pvw}{positive-velocity wing}
\newcommand{\pvws}{positive-velocity wings}

\begin{document}
\shorttitle{Multi-phase HVCs}
\shortauthors{Fox et al.}

\title{Multi-phase High-Velocity Clouds toward \he\ and \pg\footnotemark[1]}  
\footnotetext[1]{Based on observations from the NASA-CNES-CSA {\it Far
  Ultraviolet Spectroscopic Explorer} mission, operated by Johns
  Hopkins University, supported by NASA contract NAS 5-32985, and from the
  NASA/ESA {\it Hubble Space Telescope}, obtained at the Space Telescope
  Science Institute, which is operated by the Association of
  Universities for Research in Astronomy, Inc., under NASA contract
  NAS 5-26555.} 
\author{Andrew J. Fox\footnotemark[2], Bart P. Wakker, \& Blair D. Savage,}
\affil{Department of Astronomy, University of Wisconsin -
Madison, 475 North Charter St., Madison, WI 53706}
\footnotetext[2]{E-mail fox@astro.wisc.edu}

\author{Todd M. Tripp,}
\affil{Department of Astronomy, University of Massachusetts, 640
  Lederle Graduate Research Center, Amherst, MA 01003}

\author{Kenneth R. Sembach,}
\affil{Space Telescope Science Institute, 3700 San Martin Drive,
  Baltimore, MD 21218}
\and
\author{Joss Bland-Hawthorn}
\affil{Anglo-Australian Observatory, PO Box 296, Epping, NSW 2121, Australia}

\begin{abstract}
%Highly ionized high-velocity clouds (HVCs) have been detected in \os\
%absorption with the {\it Far-Ultraviolet Spectroscopic
%Explorer} (\fuse) satellite in over sixty sight lines. The origin and
%location of many of these absorbers, and their exact relation to HVCs
%detected in 21\,cm emission, is unknown.
%, particularly those detected in the form of \pvws\ -- broad components of
%\os\ absorption extending out to $>$+200\kms. 
We study the physical conditions, elemental abundances, and
kinematics of the high-velocity clouds (HVCs) along the sight lines  
toward active galaxies \he\ and \pg\ using {\it Hubble Space Telescope
  Imaging Spectrograph} 
and {\it Far Ultraviolet Spectroscopic Explorer} data. No 21-cm \hi\
emission is 
detected in these clouds, but our observations reveal multiple
components of HVC absorption in lines of \hi, \cw, \ct, \cf, \os,
\siw, \sit, and \sif\ in both directions.
%\hvo\ absorption is accompanied by coincident
%absorption in the \hi\ Lyman series and in numerous metal
%lines. The absorption in the HVCs in both sight lines contains
%both multiple components and multiple phases. 
We investigate whether
photoionization by the extragalactic background radiation or by
escaping Milky Way radiation can explain the observed ionization
pattern. We find that photoionization is a good explanation for the \cw, \ct,
\siw, and \sit\ features, but not for the \os\ or \cf\ associated with
the HVCs, suggesting that two principal phases exist: a warm
($T\approx10^4$\,K), photoionized phase and a hotter
($T=1-3\times10^5$\,K), collisionally-ionized phase; the broader line
widths of the high ions are consistent with this multi-phase hypothesis. 
%The Milky Way ionizing radiation field dominates the extragalactic background
%in a volume extending to 40\,kpc for directions $|b|<30$\degr, 100\,kpc for
%$30<|b|<60$\degr, and 180\,kpc for directions $|b|>60$\degr.
%Photoionization modelling of the clouds toward 
The warm HVCs toward \he\ have high levels of
ionization (97--99\%), and metallicities ([Z/H] between $-$0.9 and $-$0.4)
close to those in the Magellanic Stream, which 
lies eleven degrees away on the sky at similar velocities.
These HVCs may well be stripped fragments of the Stream that have been
ionized by the pervading radiation field;
%fragments of the Stream,;
they have thermal pressures that would place them close to equilibrium in a
fully ionized $10^6$\,K Galactic corona with
$n_{\mathrm{H}}=4-9\times10^{-5}$\,cm$^{-3}$ at 50\,kpc. 
The warm HVCs seen at $-$146 and 125\kms\ toward
\pg\ have [Z/H]=$-0.6\pm0.2$ and $-0.8\pm0.2$,
respectively, suggesting they are not formed from purely
Galactic material.
A mini-survey of the hot, collisionally ionized HVC components seen here
and in five other sight lines
finds that in 11/12 cases, the high ions have kinematics and ionic ratios 
that are consistent with an origin in conductive interfaces, where energy
flows into the HVCs from a hot surrounding medium and produces
\os\ and \cf-bearing boundary layers. 
However, the broad absorption wing on the \os\ profile toward \pg\ is {\it not}
completely explained by the interface scenario. This feature may be
tracing the outflow of hot gas into the Milky Way halo as part of a
Galactic fountain or wind.

\end{abstract}
\keywords{Galaxy: halo -- intergalactic medium -- ISM: clouds --
  ultraviolet: ISM} 

\section{Introduction}
Surrounding the Milky Way galaxy lies a population of diffuse gaseous
clouds that are not co-rotating with the disk. First discovered in
21-cm emission by \citet{Mu63},
% attempts to explain these so-called
%high-velocity clouds (HVCs) with a unified model have remained
%unsuccesful;
no single model has been able to simultaneously explain the
kinematics, metallicity, and emission properties of these so-called
high-velocity clouds (HVCs) \citep{WW97}. 
%Though \citet{Bl99} proposed that HVCs may
%form a population of clouds in the Local Group
The failure to find HVCs distributed throughout other galaxy groups
\citep{Pi04} together with HVC detections around M31 \citep{BT04} suggest
that HVCs are typically associated with large galaxies, rather than
the Local Group. Within this
paradigm, many HVCs \citep[e.g., Complex C,][]{Wa99} could represent material
leftover from the galaxy formation process \citep{Oo70, MB04}, or
else material stripped from nearby galaxies \citep[e.g., the Magellanic
Stream,][]{GN96}. Studying and understanding HVCs allows us to trace the
interactions and feedback between galaxies and the intergalactic
medium. HVCs regulate the rate of star formation in our
own Galaxy, by providing gaseous material from which new stars are
born. 
%Furthermore, HVCs provide an important observational constraint
%on the formation of the Milky Way, since they may trace the remnant
%material left over from the epoch of galaxy formation \citep{MB04}.

A new phase of gas in HVCs was discovered with the detections of
high-velocity \cf\ \citep{Se95, Se99} and \os\ absorption
\citep{Se00,Mu00}.
%; studying these ``highly-ionized HVCs'' provides a
%new approach for understanding HVCs in general. 
A recent survey of extragalactic sight lines with 
the {\it Far Ultraviolet Spectroscopic Explorer} (\fuse) satellite
\citep[][hereafter S03]{Se03} has found that \hvo\ absorption is
detected at $>$3$\sigma$ significance 84 times in a survey of 102
sight lines. Many of these \hvo\
features are spatially and kinematically associated with known 21-cm
emitting \hi\ structures, for example Complex C, Complex A, and the
Magellanic Stream \citep[S03,][]{Fo04}. 
%These connections can be seen
%in Figure 1, where we show the distribution of high-velocity \os\
%detections superimposed onto the high-velocity \hi\ sky, color-coded
%by velocity (S03). In cases where \hi/\os\ velocity coincidences are observed,
%the \hvo\ cannot be tracing a diffuse intergalactic phase of hot gas, but
%rather must be associated with an \hi-bearing structure. 
There are other cases where \os\ HVCs have no   
counterparts detected in \hi\ 21-cm emission. However, studies of
such highly ionized HVCs \citep{Co04, Co05, Ga05} have found absorption in
the \hi\ Lyman series and in other metal-line tracers of neutral and
weakly ionized gas at the same velocities as the high ions. The high
ion-low ion connections in HVCs are well explained by an 
arrangement where the hot gas arises in the
boundary layers between the neutral HVC and the surrounding
medium \citep{Fo04}. 
%at levels $N$(\hi)$\,\ll2\times10^{18}$\sqcm, the 21-cm
%detection limit. 
%It is unknown whether this small column of
%\hi\ represents a small neutral core, or the trace amount of neutral
%hydrogen in a predominantly ionized gas. In the first case, absorption line
%observations would be showing that \hi\ HVCs exist down to much
%lower column densities than known before.
%Indeed, a recent study of the
%21-cm $N$(\hi) distribution function by \citet{Wk04} using data from
%\citet{Mu95} shows the distribution to be rising to lower $N$(\hi) at
%the lowest measured $N$(\hi) bin.

%Based on comparing observational data with the 
%prediction of models, S03, \citet{Co04,Co05}, \citet{Fo04} and \citet{Ga05}
%have offered various explanations for the 
%physical mechanism producing \os, including bow shocks, conductive
%interfaces, and turbulent mixing layers, all of which involve the
%interaction of the cloud with a surrounding medium. 

A separate class of \hvo\ absorber exists in the form of
broad, shallow absorption wings at positive velocities. Unlike many of
the discrete high velocity \os\ absorption features seen 
along complete paths through the Galactic halo and corona, the \os\
\pvws\ manifest themselves as continuous absorption troughs extending from
velocities indicative of Galactic halo gas ($|v_{LSR}|<100$\kms) out to
velocities of several hundred \kms. A majority of \pvws\ (17/21) are
found in the 
Northern Galactic hemisphere, at $l$=180-300\degr and $b>$30\degr, even
though the sample of sight lines is spread fairly evenly around the
sky (S03). Such a trend may be related to the general enhancement of
\os\ in the 
Northern hemisphere of the Galactic halo \citep{Sa03}, and to the opening
of the Local Bubble in this direction \citep{We99}. 
%Considerations
%of the presence of other hot interstellar structures
%in wing directions led \citet{Se01a} and \citet{Sa05a} to suggest that
%the wings may be tracing an outflow of hot gas from the Galactic disk. This
%explanation implies that the wings are produced by a different
%physical process than the \hvo\ components, rather than simply
%representing unresolved discrete absorbers. However, detailed studies
%of the physical conditions in the wings do not yet exist.

%To explore the differences between \os\ wings and components, 
In this paper we present high-resolution Space Telescope Imaging
Spectrograph (STIS) observations complemented with \fuse\ data to 
explore the \hva\ in one direction showing \hvo\ components (\he) and 
one showing a \hvo\ wing (\pg). Basic 
properties of the two sight lines under study are found in Table 1. 
%No high-velocity 21-cm emission is detected in these directions, though
%there are nearby 
%high-velocity \hi\ structures, as can be seen in Figure 1. 
%In order
%to understand the origins of highly ionized high-velocity gas we look
%for relationships between absorption in highly ionized species 
%and absorption in neutral and weakly ionized species. We also search
%the \hi\ Lyman series absorption lines to look for low column density \hi\
%neutral counterparts to the \os\ HVCs. We use photoionization modeling
%to explore the physical conditions and abundances in the neutral HVCs,
%providing useful information on possible cloud locations.

The line of sight to Seyfert~1 galaxy \he\ ($l=253.94\degr, b=-65.78\degr$,
$z_{em}=0.495$) lies $10.8\degr$ from the
$N$(\hi)$=2.0\times10^{18}$\sqcm\ (5$\sigma$) contour of \hi\ 21-cm
emission from the Magellanic Stream (Figure 1, left panel).
% In the left panel of
%Figure 1 we display a map of high-positive-velocity 21-cm emission from this
%direction, showing the location of the \he\ direction with respect to
%the body of the Stream. 
The Magellanic Stream is a broad filament of 
material covering $\sim$1000 square degrees in an arc 
extending from $l\approx90\degr, b\approx-45\degr$, through the South 
Galactic pole and around to positive latitudes \citep{Pu03a,
  Br04}. The Stream is believed to be gas stripped out of
the Magellanic Clouds by %both ram pressure and 
tidal forces as the Clouds orbit the Milky Way \citep{GN96}. 
Six sight lines passing through the Magellanic Stream have 
\os\ detections at similar velocities to the \hi\ (Figure 15 in S03),
suggesting that the Stream contains (or is surrounded by) a highly
ionized component. 

\pg\ is a Seyfert~1 galaxy at $z_{em}=0.239$ lying at $l$=179.79\degr,
$b$=+51.71\degr. Various nearby \hi\ structures can be seen in Figure 1
(right panel).
%, which shows a map of the high-negative-velocity 21-cm
%emission in the vicinity of the sight line. 
This direction has the advantage of being very close the anticenter
direction ($l$=180\degr), perpendicular to the direction of LSR
motion, so the effects of Galactic rotation on 
absorption line profiles are removed. \pg\ is 
therefore ideally situated for use as a backlight to study %motions of
%interstellar clouds that are unrelated to Galactic rotation, such as
the inflow and outflow of gaseous material. 

The structure of this paper is as follows. In \S2 we describe the
observations and data reduction. In \S3 our spectroscopic
measurement and analysis techniques are discussed. In \S4 and \S5 we
discuss our observations of the high-velocity absorption
systems seen toward \he\ and \pg, respectively. We discuss
photoionization modeling of the HVCs in \S6, and collisional
ionization modeling of the highly ionized species in \S7. Section 8 is
devoted to understanding the origin of the \os\ wing absorption seen
toward \pg. The principal results of our study are summarized in \S9.

\section{Observations and Data Reduction}

%The STIS instrument has
%several echelle modes for high-resolution ultraviolet spectroscopy.
\he\ and \pg\ were each observed on several occasions with the E140M
echelle mode of the STIS instrument \citep{Wo98, Ki98}, which provides
spectra at 7\kms\ resolution (FWHM) in 
the range 1150--1700\AA, with a few small gaps between orders at
$\lambda\!>$1600\AA. Details of these observations are given in Table
2. The data were reduced as described in \citet{Tr01} including the
two-dimensional scattered light correction developed by the STIS
Instrument Definition Team \citep{LB97, Bo98}. In general, the
wavelength calibration of STIS data is 
excellent; \citet{KQ03} report that the relative
wavelength calibration is accurate to 0.25--0.5 pixels across the
entire spectrum, and the absolute wavelength calibration is accurate
to $\sim$1 pixel (3.2\kms). 
%In a few cases slightly larger wavelength
%calibration errors have been noted \citep[e.g.][]{Tr04}, but
%these occur infrequently. 
The overall velocity uncertainty in our data
was verified by
cross-checking the centers of Galactic metal-line absorption in \oi\
$\lambda1302.169$, \few\ $\lambda1608.451$, \siw\ $\lambda1304.370$,
\siw\ $\lambda1526.707$, \ion{Ni}{2} $\lambda1317.217$, \ion{S}{2}
$\lambda1253.811$, and \ion{S}{2} $\lambda1250.584$. 

The \fuse\ satellite \citep{Mo00, Sa00} provided the far-UV
spectra of \he\ and \pg\ (Table 2).
%consists of four co-aligned
%Rowland-circle spectrographs, producing medium-resolution far-UV
%spectra between 905 and 1187\,\AA. Two of the
%spectrograph channels are coated with Al+LiF for optimum performance
%above 1000\,\AA, and two are coated with SiC for higher reflectivity at
%wavelengths below 1000\,\AA. The resolution (FWHM) in the SiC channels is
%$\approx25$\kms, and in the LiF channels, $\approx20$\kms. Our \fuse\
%observations of \he\ and \pg\ 
%(Table 2) were reduced with the data reduction pipeline CALFUSE
%(v2.1.6 or above).
%, which provides a series of reduction steps to
%produce flux-calibrated, wavelength-calibrated spectra. These steps
%consist of screening out data taken when the satellite passed through the
%South Atlantic Anomaly and pointed through the Earth's limb,
%correcting for detector drift, geometric distortions, grating rotation
%and satellite motion, subtracting the sky background, calibrating the
%wavelength scale, and converting the count rate into a flux. 
For wavelengths below 1000\,\AA, we use data from the SiC2A detector
segment only. We did not combine with data from the SiC1B
detector segment data because this detector has lower sensitivity and
resolution. For 
wavelengths between 1000 and 1187\,\AA, we use coadded 
LiF1 and LiF2 data, since the resolution difference between these two
channels is small. The data were reduced with the
data reduction pipeline CALFUSE (v2.1.6 or v2.4.0). Even after the
application of the pipeline, velocity shifts have to be
determined for each exposure and  
each detector segment before coaddition of the spectra to obtain
accurate wavelength solutions \citep[see \S3.4 in][]{Wa03}.
We measured the average velocity centroid of all the unsaturated neutral ISM lines clearly
detected in each \fuse\ 
segment (\siw\ $\lambda1020.699$, \ion{Ar}{1} $\lambda1048.220$, \ion{N}{1}
$\lambda1134.165, 1134.415, 1134.980$, and numerous
H$_2$, \few, and \oi\ lines), and then determined the offset from the
centers of neutral ISM lines seen in the STIS data. This ``bootstrapping''
technique allows the kinematics of absorption line profiles observed
with \fuse\ and STIS to be reliably compared.  All velocities quoted in
this paper are referenced to the Local Standard of Rest (LSR). Note
that no velocity 
difference was found between the metal lines and the H$_2$ lines. The
offsets for the LiF channels are typically $\approx10$\kms\ but in the
SiC channels they can be as high as 
$\approx40$\kms. We estimate a residual 1$\sigma$ zero-point error of 5\kms\ 
after this process has been applied, since 5\kms\ is the dispersion of
post-correction velocity centroid measurements of neutral lines in the \fuse\
data. This velocity error is propagated in calculating the systematic 
errors on our absorption line measurements. 

Continua were fitted to both the \fuse\ and STIS data in regions
approximately 1000\kms\ wide around each absorption line of interest, using
regions of the spectra judged to be free from absorption. In almost
all cases a linear continuum could be fitted, since our target active
galaxies have flat UV spectra generated by non-thermal emission
processes. 

%Raw \fuse\ spectra are highly oversampled, with ten 2\kms\
%pixels per 20\kms\ resolution element. Therefore 
All \fuse\ data were rebinned by five pixels to provide optimal
sampling, for both display and measurement.
%; the measurements we make
%and the data we show use this rebinned \fuse\ data. 
The raw STIS spectra are already optimally sampled, 
%with 3.2\kms\ pixels and a 6.7\kms\ resolution element. 
so we measure the absorption using the unbinned data, but
%In theory, this data should not be
%resampled, so our measurements are made on the raw data. However,
since the STIS spectra for both \he\ and \pg\ are very noisy,
we use 2-pixel rebinning for display\footnotemark.
\footnotetext{Exception: in one case, where we need the highest
  possible resolution to emphasize the component structure, we show
  the fully sampled spectra (\S4.3).}  
%The measurements of the STIS absorpion lines were made on the
%unbinned, optimally sampled data.

For each of our two studied sight lines we display \hi\ 21-cm
emission spectra, to emphasize the low $N$(\hi) of the highly ionized
HVCs under study. For the \he\ direction these data were obtained and
provided by R. Morras, using the Villa Elisa telescope, which has a 
beam size of 34$\arcmin$; some residual uncertainty in the baseline
fit is present in the range $\pm100$\kms. For 
\pg, we use spectra from the Effelsberg
telescope \citep{Wa01} with a 9$\arcmin$.7 beam. Both data sets have
an approximate velocity resolution of 1\kms, and have been corrected for side
lobe contamination. 

\section{Spectroscopic Measurements}

We identified the high-velocity
component structure in each sight line using the high-resolution STIS
observations of metal-line absorption, particularly in the \cw\
$\lambda1334.532$, \siw\ $\lambda1260.422$, \siw\ $\lambda1193.290$,
and \sit\ $\lambda1206.500$ lines. We then
estimated the velocity range covered by each absorption component, and
measured the equivalent width of the absorption between these limits
in all spectral lines present in the data. We converted the
measured optical depth in the line as a function of velocity into
apparent column densities, $N_a(v)$, using the relation
$N_a(v)=3.768\times10^{14}(f\lambda)^{-1}\tau_a(v)$
\,ions\sqcm(km\,s$^{-1}$)$^{-1}$, where the 
apparent optical depth (AOD) is given by $\tau_a(v)={\rm ln}[F_c(v)/F(v)]$,
$f$ is the oscillator strength of the transition \citep[taken
  from][]{Mn03}, $\lambda$ is the transition wavelength in Angstroms,
and $F(v)$ and $F_c(v)$ are the observed line and continuum fluxes at
velocity $v$, respectively \citep{SS91}. Integrating $N_a(v)$ between
two chosen velocity limits determines
the total column density in that velocity range, and the optical
depth-weighted average values of $v$ and $\sqrt(v-\bar{v})^2$, i.e.
the first and second moments of $\tau_a(v)$, give
measurements of the line center and line width \citep{SS92}. 

The AOD technique requires no prior knowledge of the
component structure.
%The apparent column density method is our primary technique for
%measuring the strength of high-velocity absorption in 
%various species, 
It will produce good results as long as the line
profiles are resolved and not saturated. STIS/E140M, \fuse/LiF,
and \fuse/SiC can resolve lines with $b>$4, 12, and 15\kms, respectively. 
\citet{Wa03} show a comparison of the apparent column density of \os\
from both members of the doublet toward \he\ and \pg\ in the velocity range
$-$400 to 400\kms; this data shows that no
unresolved saturation affects the zero-velocity (Milky Way disk and
halo) \os\ absorption. By inference, the weaker high-velocity \os\ absorption,
of interest here, is also likely to be unsaturated.
The \os\ lines are likely to be resolved since the 
high temperatures of the regions where \os\ is found result in 
lines with thermal widths $b>18$\kms.
%Among the lines observed with 
%\fuse, \os\ is likely to be resolved, since the large thermal
%broadening arising from the high temperature of the locations where it
%is typically found produces $b>18$\kms. 
However, \fuse\ absorption line profiles of the \hi\ Lyman
lines may be unresolved, making saturation hard to detect. 
%In these cases, the apparent
%optical depth method will not return reliable estimates of the column
%density in those species. 
%We test for saturation for
%looking for trends in which $N_a$(\hi) increases when integrating over
%the same velocity range in progressively weaker Lyman
%series absorption lines. 
As one measures progressively weaker Lyman series absorption lines, the
$f$-values of the transitions decrease, and eventually the lines will
become unsaturated. At this point, consecutive Lyman lines will give the
same value for $N_a$; this occured for the HVCs along our two sight lines with the
\hi\ $\lambda$ 920.963, 919.351, 918.129, and 917.181 lines.
To check the column densities derived from AOD integrations of these
lines, we construct a 
Voigt profile model and compare with the data. For the
low-velocity components, we take the parameters from the fits to the
21-cm profiles in our two target directions
\citep{Wa03}; for the high-velocity components,
we take the velocities, widths, and column
densities defined by the AOD integrations. All components are convolved
with a Gaussian line spread function (with FWHM 25\kms\ for the
\fuse/SiC channels). Our model also reproduces the \ion{D}{1} and \oi\
lines when D/H$=2\times10^{-5}$ and O/H is solar,
assuming they have the same width as the \hi\ lines (i.e., non-thermal broadening dominates).
% ($4.6\times10^{-4})$
Small adjustments ($\approx5$\kms) are needed to the {\it widths} of
each \hi\ component to produce a good fit\footnotemark, but
the AOD {\it column densities} reproduce the depth of absorption well.
\footnotetext{For \pg, we found that in addition to the HVCs measured
  here, two extra components were needed to reproduce the Lyman series
  and \oi\ absorption line profiles. A component at $-$110\kms, with
  $N$(\hi)$=1-2\times10^{17}$ is needed. This component is also seen
  in the negative-velocity wings of \siw\ $\lambda1260.422$ and \sit\
  $\lambda1206.500$, and in \hi\ 21-cm emission at this velocity less
  than 2\degr\ away, in HVC MIII (see Figure 1). A second component at
  +55\kms\ with 
  $N$(\hi)$=1-2\times10^{17}$\sqcm\ is needed to fill in the \hi\
  absorption between the zero-velocity component and the
  positive-velocity HVC. Neither of these components changes the \hi\
  column density  of the HVCs under study in this paper. }
We thus have confidence in the values of $N$(\hi) derived from the
AOD method. A full Voigt profile {\it fit} of the \hi\ Lyman series
absorption lines, involving minimizing $\chi^2$ 
for a grid of different values of log\,$N$(\hi) and $b$(\hi), was
attempted, but did not produce reliable results given the overlapping
component structure, poor S/N, numerous blends, and degeneracy between
$N$ and $b$. 
%\hi\ 21-cm
%emission line measurements are not
%sensitive enough to derive $N$(\hi) in the weak components, and they 
%also suffer from a larger beam size that cannot be reliably compared to
%absorption lines measured over an effectively infinitesimal beam. 
%We therefore
%decided that AOD integrations on the weakest measured Lyman series
%absorption lines represented the best method for measuring the \hi\
%column density. 

To check for the possibility of saturation
in the strong metal lines (\cw, \ct, \siw, \sit), Voigt profile models 
were constructed in the same
manner as for the \hi\ lines (using the column densities from the
$N_a(v)$ integrations) and compared with the observations. We
investigated how saturated the observed profiles could be by
increasing the model component column densities and noting
whether the quality of the fit decreases. For the cases of the
\ct\ and \sit\ lines toward \he, saturation is clearly
possible in some HVC components, resulting in large positive
errors (as large as 1\,dex) in log\,$N$.

The full absorption line measurements are presented on a
component-by-component basis for the \he\ system in Table 3, and
for the \pg\ system in 
Table 4. We list the component velocity, width, and column
density measurements returned by the AOD method, the
equivalent width, and the observed S/N near the line, measured per
resolution element in the continuum.
%\footnotemark.
%\footnotetext{The noise level is derived by measuring the rms
%  dispersion of the data points around the fitted continuum.}
%Within each table, the results are presented in order of
%increasing atomic number of the element causing the transition, and
%increasing ionization stage within each element. Where
%multiple lines from the same ion are present, we give the results of
%the strongest line first and continue down in oscillator strength.
When no absorption is detected in a particular feature, we calculate
a 3$\sigma$ upper limit for the ionic column density using the
error on the equivalent width measurement, converted to a
column density limit assuming a linear curve-of-growth\footnotemark.
\footnotetext{The relation used is
  $N=1.13\times10^{17}W_{\lambda}/\lambda^2f$, 
  where $N$ is in\sqcm, $W_{\lambda}$ is the 3$\sigma$
  equivalent width limit in m\AA, and $\lambda$ is in \AA.}

Two sets of errors are quoted for each measurement of equivalent width
and column density. The first is a quadrature sum of statistical
uncertainty in the count rate (Poisson noise) and the continuum 
placement uncertainty. The second is the systematic error, found from 
a quadrature addition of fixed pattern noise (f.p.n.) in the detectors with
the velocity scale errors. We estimated the f.p.n. per
unbinned pixel to be 1\% for STIS detectors, 2.5\% for \fuse\
observations of \he, and 10\% for \fuse\ observations of \pg. 
The f.p.n. is lower in the \he\ \fuse\ dataset due to the use
of focal-plane (FP) splits for some observations, which place the spectra in
different positions on the detector. These errors imply that
any feature weaker than 1.8\,m\AA\ (6.3\,m\AA) could be caused by 
f.p.n. in the \he\ (\pg) \fuse\ data. The velocity error ($\pm$5\kms)
allows for the intrinsic uncertainty in the zero point of the post-calibration
velocity scale, and the uncertainty in the choice of 
velocities used to define a particular high-velocity component. The latter
uncertainty is important since the different components of 
high-velocity absorption are not always distinct from each other and
placing the division can be difficult. We calculate the effect of this
error by changing the velocity extrema of each component by $\pm5$\kms\ and
observing the change in the equivalent width and in $N_a$.

For each sight line we have prepared three plots showing the
absorption line data (Figures 2, 3, and 4
for \he\ and Figures 5, 6, and 7 for \pg).
First we display (Figs. 2 and 5) the fully-reduced spectra for all absorption
lines in which high-velocity absorption is detected, using gray
shading to identify the gas in the high-velocity 
components. We also include \hi\ 21-cm emission line spectra of
our two sight lines in the upper left panels. 
%In each
%panel the continuum is marked with a dashed line and blends have been
%identified with tick marks. 
In Figures 3 and 6 we include stacks of \hi\ Lyman series absorption
profiles. These 
figures show the presence of high-velocity \hi\ absorption in the
Lyman series from Ly$\gamma$ at 972.537\,\AA\ ($n=1\!\to\!4$) down to
Ly$\mu$ at 917.181\,\AA\ ($n=1\!\to\!13$). At wavelengths shortward of
Ly$\mu$ the flux goes to zero. Ly$\alpha$ is too broad and
Ly$\beta$ is too polluted by geocoronal emission (\oi\ airglow) to show the
high-velocity component structure. Geocoronal emission 
in the stronger Lyman lines tends to ``fill in'' absorption 
in the approximate range $-$100 to 100\kms, so we display night-only
data for the \hi\ lines to minimize this effect; at high velocities
the geocoronal emission becomes insignificant.
Our final method of displaying the data is to compare the apparent
column density profiles of particular lines (Figures
4 and 7). 

\subsection{Blends toward \he}
An IGM absorption line system at $z=0.34028$ is seen in
Ly$\alpha$, Ly$\beta$, \ct, \ion{O}{4}, and \os\
\citep{Le05}. Absorption at this 
redshift by \ion{Ne}{8} $\lambda770.409$ would blend with \hva\ from
the strong line of \os\ at 1031.926\,\AA\ in the range
150--210\kms. We used the redshifted absorption data for the weak line in 
the \ion{Ne}{8} doublet ($\lambda780.324$) to assess the strength of
the blend; the weak line shows no absorption at the same redshift, with
$W_{\lambda}$(\ion{Ne}{8})$<$9\,m\AA\ (3$\sigma$) integrated over
60\kms\ (the width of the \os\ lines in the IGM system). Therefore the strength
of the \ion{Ne}{8} $\lambda770.409$ blend must be $<$18\,m\AA\
(3$\sigma$), since $f_{770}/f_{780}=2$. The
total absorption in the range 160--230\kms\ in the frame of the \os\
line has $W_{\lambda}=45\pm5\pm5$\,m\AA, more than can be accounted
for by the blend, so we proceed by treating this as a true detection of \os. 
%The
%total absorption in the range 160--230\kms\ in the frame of the \os\
%line has $W_{\lambda}=45\pm5\pm5$\,m\AA, implying
%$W_{\lambda}$(\os)$>$19\,m\AA\ (3$\sigma$) when the blend is
%considered and the errors are added in quadrature.
% For the two components in the range 160--185 and 185--230\kms, we
% therefore adopt $W_{\lambda}$(\os)$>$9\,m\AA\ in each.

%Further information on the likely strength of this blend is
%provided by measurement of \os\ and \ion{Ne}{8} in another
%intergalactic system toward \he, at $z=0.20701$ \citep[see][]{Sa05b}. If
%$W_{\lambda}$(\ion{Ne}{8} $\lambda770.409$)/$W_{\lambda}$(\os\
%$\lambda1031.926$) = 0.2 in the $z=0.34028$ system, as in the
%$z=0.20701$ system, then we would predict
%$W_{\lambda}$(\ion{Ne}{8} $\lambda770.409$)$\approx 13$\,m\AA\ in the
%offending system, suggesting that some degree of contamination
%is likely. 

Absorption from the $J=4$ rotational level of molecular hydrogen at
1032.156\,\AA\ blends with \hva\ 
from \os\ $\lambda1031.926$, and absorption from the $J=1$
level at 1038.156\,\AA\ blends with \hva\ from
the  \os\ $\lambda1037.617$, {\it if} the H$_2$ column
densities are high enough. We employed a modeling process to decontaminate the
data from blending by H$_2$ lines. We took all H$_2$ lines in the
\fuse\ data with no major contamination or continuum placement
uncertainties, and measured their equivalent widths. We
then performed a curve-of-growth analysis for each rotational ($J$) level, and
derived log\,$N$(H$_2$) = 13.72$\pm$0.15, 13.99$\pm$0.23, $<$13.89,
$<$13.89, and $<$13.83 in levels
$J=0,1,2,3,$ and 4 respectively, corresponding to
log\,$N$(H$_2$)$\approx$14.34 in total. This model is included in the
continuum fits for contaminated \fuse\ absorption lines in Figure 2. 
%Given this information, the H$_2$ $\lambda1032.156$ 
%line at 125\kms\ in the frame of \os\ $\lambda1031.926$ is predicted
%to be of negligible strength (central line depth
%$<$0.01). However, the 1038.156\,\AA\ line should have a depth
%of 0.10; we find absorption in the \os\ $\lambda1037.617$ profile
%close to the expected velocity of the H$_2$ blend with a depth
%$\approx$0.2. 
Examination of the \os\ $\lambda1037.617$ panel shows significant
contamination of the high-velocity absorption by H$_2$
$\lambda1038.156$, and so we choose to base all measurements of \hvo\
on the strong \os\ line. 
%Given the \ion{Ne}{8} contamination problems for the
%strong \os\ line, we can only proceed with a lower limit to the \os\
%absorption in the high-velocity components above 150\kms.

Contamination from redshifted \os\ $\lambda1037.617$ absorption associated with
the AGN itself ($z=0.49246$) blends with low- and high-velocity
\cf\ absorption in the 1548.195\AA\ line. This associated absorber has
a multiple-component structure and prevents a reliable measurement
of high-velocity \cf\ $\lambda1548.195$ absorption. We therefore
concentrate on the weak line (1550.770\AA) for measurements of
high-velocity \cf\ absorption in this direction.

\subsection{Blends toward \pg}
Our measurement of the foreground H$_2$ absorption in the \pg\
direction finds log\,$N$(H$_2$) = 13.90, 14.40, 14.00, 14.12, $<$14.09 in
levels $J=0,1,2,3,$ and 4, giving a total log\,$N$(H$_2$) = 14.75.
As was the case for \he, H$_2$ $\lambda1032.156$ at 125\kms\ in the
frame of \os\ $\lambda1031.926$ is predicted to be of negligible
stength, but the $J=1$ H$_2$ line at 1038.156\,\AA\ (central line depth 0.27)
blends with \hva\ from the weak line of \os\ at 1037.617\,\AA\ (See
Figure 5). We therefore adopt measurements of the strong line of \os\ only.
 
There are few IGM systems near our lines of interest in the \pg\
sight line. See \citet{Sa02} for an investigation of the IGM
absorption in this direction.

\section{High-Velocity Absorption toward \he}
Our \fuse\ and \hst/STIS observations of \he\ reveal
high-velocity absorption in lines of \hi, \cw, \ct, \cf, \os, \siw,
\sit, and \sif\ (Figure 2). \few\ is detected at low significance.
No absorption is seen in \ion{N}{1},
\ion{N}{2}, \oi, \ion{S}{2}, \ion{S}{3}, \ion{P}{2}, \ion{Ar}{1}, and
\ion{Fe}{3}. Multi-component structure is clearly seen
in the HVCs in the \siw, \sit, \cw, and \cf\ profiles observed at 7\kms\
resolution. Based on component fits to these lines, we identify four
positive-velocity components centered at 99, 148, 175, and
193\kms, covering the ranges 80--125, 125--160, 160--185, and
185--230\kms, and shown in Figure 4 with the colors purple, blue,
green, and yellow 
respectively. Absorption in the 99\kms\ component is distinct from the
Galactic absorption centered at $-$10\kms, so we treat this component
as an HVC, even though the blueshifted parts of this component lie outside
the formal definition of high-velocity ($|v_{LSR}|>100$\kms). 
%The column densities of each species detected in each component are
%presented in Table 3, together with upper limits for \ion{N}{1},
%\ion{N}{2}, \oi, \ion{S}{2}, \ion{S}{3}, \ion{P}{2}, \ion{Ar}{1}, and
%\ion{Fe}{3}, all of which 
%show no detectable absorption in any of the HVC components. 
%The
%measurements are made on a component-by-component basis, rather than
%integrated over the whole range of HVC absorption, since each
%component may have unique ionization conditions. 

Inspection of Figure 4 reveals the HVC component structure to be consistent
among lines of different ionization states. We show the STIS data at full
resolution here so that no kinematic information is lost. 
%For this
%plot only, we use
%the combined day+night data for \ct\ to maximize the S/N. 
Tracers of neutral (\cw, \siw), weakly ionized (\sit), and highly ionized gas
(\cf, \sif, \os) show absorption over the same velocity 
range, and with a similar component structure. Among the high ions,
the 99 (purple), 175 (green), and 193\kms\ (yellow) components are
clearest in \os, \cf, and \sif, respectively.  
%, seen in \cw\ and \siw, show corresponding
%highly ionized components, 
%albeit in different species in each case: 
%the 99\kms\ component is clearest in \os, and
%the 193\kms\ component is 
%clearest in \sif. 
This indicates that the high-ion/high-ion ratios differ between components.
There are further differences
between components in the low-ion/high-ion ratios; for example,
$N$(\siw)/$N$(\sif) changes from $>2.0$ at 145\kms\ to $1.0$ at 193\kms.

The unblended \os\ absorption in the 99\kms\ component is
broader than the corresponding lower ionization component; we measure
$b$(\os)=33$\pm$2\kms\ and $b$(\cw)=16$\pm$3\kms. 
The fact that \hva\ is seen over a similar velocity extent in
both neutral, weakly ionized, and highly ionized species indicates
that common structures contain all these ions. However, the line
widths imply that the neutral and highly ionized species are not
co-spatial, and so {\it multiple phases} are required in each component. 

%\few\ $\lambda1144.938$ absorption in the 148\kms\
%component is detected with $W_{\lambda}=11.2\pm4.6$, corresponding to
%log\,$N=13.01^{+0.14}_{-0.21}$ and a formal significance of only
%2.4$\sigma$. However, we verified the presence of this \few\ feature
%in both \fuse\ channel segments that cover wavelengths longward of
%1100\,\AA\ (LiF1B and LiF2A). 
%Furthermore, the component is at essentially the same velocity (145\kms)
%as the component seen in other tracers of neutral gas (\cw\ and \siw),
%so we proceed for now by treating \few\ as being detected.
%The other \few\ line in our data set at 1608.451\,\AA\ shows
%no corresponding \hva, but the 3$\sigma$ upper limit set on log\,$N$(\few)
%using this line is 13.50, consistent with the value obtained from the
%1144.938\,\AA\ line. The S/N in the STIS data near this line (=6) is too low
%for a better constraint. 

The \hi\ Lyman lines (Figure 3) show strong \hva\ extending out to 230\kms, the
same extreme velocity as is seen in the \os\ profile.
% Our model Voigt profile fit is shown as the red line. 
%Despite our use of night-only data, residual airglow emission is still present
%near $v_{LSR}=0$\kms\ in the stronger lines.
%, and so the fits appear poor in the low-velocity regions. 
We note that the \hi\ column densities in the four HVCs toward
\he\ all lie between $10^{16.2}$ and $10^{16.7}$.
%, prompting us to
%wonder whether some physical process fine-tunes the cloud size to this
%particular range. 
Since \hi\ clouds become optically thick to hydrogen 
ionizing radiation ($\lambda<912$\,\AA) at
log\,$N$(\hi)$\,\approx17.2$, these HVCs are optically
thin. Consequently, there 
should be little shielding of the HVC core, and we expect the ionizing
radiation field to be relatively constant within the clouds.

\section{High-Velocity Absorption toward \pg}

%The \pg\ spectrum shows a distinct \pvw\ in the \os\ absorption line
%profile (Fig. 5). Our STIS observations allow us to investigate the
%nature of the \pg\ wing by searching for corresponding absorption in
%other species. The \pg\ sight line is 13\degr from the direction toward
%the BL Lac object Mrk~421 ($l$=179.83\degr, $b$=65.03\degr), where
%another \os\ \pvw\ is detected \citep{Sa05a}.
%\subsection{Likelihood of Saturation}
%The comparison of apparent column density profiles from each member of
%the \os\ doublet in the \pg\ direction \citep{Wa03} shows no
%unresolved saturation in the zero-velocity (Milky Way disk and
%halo) absorption, implying the weaker high-velocity absorption
%should also be unsaturated. The $N_a$(\os) measurements in the \hva\ are
%therefore good estimates of the true column density.

\fuse\ and STIS data of \pg\ reveal \hva\ in \hi, \cw, \ct, \cf,
\sit, \sif, and \os\ in the
range 80--200\kms, together with a narrow $-$146\kms\ HVC seen in the
lines of \cw, \siw, and \sit\ (Figure 5, Table 4). 
%\hi\ absorption in the HVCs is clear 
%in all ten Lyman series absorption lines shown in Figure 6. 
%The intercomparison between lines of different species in the
%positive-velocity HVC can be seen more
%clearly in the $N_a(v)$ profiles displayed in Figure 7, where we note
%several important
%differences between the profiles of different species. 

In the positive-velocity HVC, a single, narrow ($b=10$\kms) neutral 
component at 124\kms\ is clearly
seen in the \cw\ profile (Figure 7, blue shading), and at moderate
significance (3.0$\sigma$) in the \oi\ profile, though no
coincident \siw\ absorption is detected at significant levels. 
The \ct\ and \sit\ profiles (green), tracing gas of intermediate 
ionization level, also show a component centered near 125\kms, though
the absorption in these species is much broader ($b=40$\kms); 
the \ct\ profile also contains a wing extending out to 200\kms. 

The highly ionized species \cf\ and \sif\ (red) show weak absorption components
centered near 150\kms\ at detection significances of 2.7 and
2.3$\sigma$, respectively, displaced in velocity from the \ct\
centroid by $\approx$25\kms. 
The \os\ profile in this direction shows a distinctive \pvw\ extending
from 75\kms\ out to 200\kms. The wing is
qualitatively different from the absorption profiles in neutral
species, weakly ionized species, and the highly 
ionized species \cf\ and \sif. 
%The difference between the \os\ and
%\cf\ profiles is particularly significant, since in only one other
%\os\ wing sight line \citep[3C273.0;][]{Se01a} do profiles of the
%accompanying \cf\ and \sif\ absorption exist. 
Interestingly, there
is more \os\ absorption at velocities where \hi\ is strongest
(80--125\kms) than where \cf\ is strongest (135--200\kms); the
$N$(\cf)/$N$(\os) ratio therefore changes with velocity over the range of the
\hva. 

%The similarity between the \os\ profile and the \hi\ profile is shown 
%using yellow coloring.
%, although we note that the 
%absorption in \hi\ recovers to the continuum by 175\kms, whereas
%absorption in \os\ continues out to 200\kms. The \os\ wing is therefore
%different from the \os\ components toward \he\ (\S4.3), because in
%that case the \hi\ and \os\ extend to the same maximum velocity.
% This very high level of ionization is a characteristic signature of \pvws.

The negative-velocity HVC at $-$146\kms\ is seen in
%log~$N_a$(\cw)=13.57, log~$N_a$(\siw)=12.57, log~$N_a$(\sit)=12.55,
%and log~$N_a$(\sif)=12.39. All four lines are 
lines of \cw, \siw, \sit, and \sif, all of which show $b$-values in
the range 12--15\kms. Although the \hi\ absorption in this
component appears to be blended with low-velocity
absorption, we conclude high-velocity \hi\ is present since our Voigt
model without the $-$146\kms\ component clearly does not fit the data
(Figure 6).
The $-$146\kms\ HVC has the interesting
property of no associated \os\ or \cf\ absorption. 
%We measure log\,$N_a$(\os)$<13.67$ and log\,$N_a$(\cf)$<13.19$ at 3$\sigma$
%significance between $-190$ and $-110$\kms. 

\section{Photoionization Modelling: the Cloud Cores}

%Our approach to understanding the origin of the HVC
%systems toward \he\ and \pg\ is to investigate the mechanism
%responsible for maintaining the observed level of ionization in the
%clouds. If determined, this mechanism can provide clues on the HVC
%location, which remains the key uncertainty.
We used the 1-dimensional plane-parallel
photoionization code CLOUDY \citep[v94.00;][]{Fe98} to model the HVCs
as uniform density slabs exposed to an ionizing radiation field.
The purpose of this exercise is to assess whether
photoionization is a viable ionization mechanism, and if so, to provide
information on the physical conditions and elemental abundances in the
clouds. %We assume the HVCs to have uniform density.
%Our models are critically dependent on the radiation field incident upon the
%clouds. 
We consider two incident radiation fields: the extragalactic
background (EGB) and the radiation field escaping from the Milky Way (MW).

%test whether the 
%observed ionic column densities in each HVC component can be explained by
%one value of the ionization parameter $U$ and metallicity [$Z$/H], where
%and 

\subsection{Extragalactic Background Radiation}

The extragalactic background (EGB) radiation contains enough ionizing
photons to maintain the levels of ionization seen in many HVCs
\citep{Se01a, Tr03, Co04, Co05, Ga05}. The EGB field we use is taken from
\citet{HM96}, who calculate the mean intensity $J_\nu$ as a
function of wavelength at various redshifts by following the radiative
transfer of a model QSO UV spectrum through the IGM  (see their Figure 5). The
EGB field at $z=0$ has a mean intensity at the Lyman Limit (912\,\AA)
$J_\nu=10^{-23}$\,erg\sqcm\,s$^{-1}$Hz$^{-1}$sr$^{-1}$, and is shown
as the blue line on Figure 8.

A CLOUDY calculation requires a geometry in which the radiation
field is plane-parallel and incident from one side of the cloud. For
the EGB, the true field is isotropic, 
but since the HVCs under study are optically thin to Lyman continuum
radiation, the {\it direction} of incident ionizing photons is
unimportant. We therefore collapse the EGB field into a
one-dimensional flux with the same energy density as the isotropic
field ($F_\nu=4\pi J_\nu$). When one divides $F_\nu$ by the photon
energy $h\nu$ and integrates from 0 to 912\,\AA, the 
EGB field at $z=0$ is found to contain 10$^4$ hydrogen-ionizing (hard)
photons\sqcm\,s$^{-1}$.
%; integrating from 912 to 2460\,\AA\ finds a
%total of 6.7$\times10^4$ non-hydrogen-ionizing (soft)
%photons\sqcm\,s$^{-1}$ in the EGB field.

\subsection{Escaping Milky Way Radiation}

%For HVCs close to the Milky Way, escaping Galactic radiation
%will affect the ionization structure of the HVCs. 
To determine the flux of escaping Galactic radiation, we first
estimate the ionizing spectrum in
the disk, and then consider what fraction of photons escape into the halo. 
The basic theory is outlined in \citet{BM99} although we
now present an extended model. First, the 90-912\,\AA\ region is
dominated by O-B stars and these are confined to spiral arms
\citep[see][]{BM02}.
The unextinguished O-B star radiation field is normalized
to the distribution of O stars near the solar circle compiled by
\citet{Va96}. The mean opacity in the 90-912\,\AA\ window is taken to
be $\tau$=2.8, implying 6\% of the hard O-B star photons escape into the
halo normal to the disk. This escape fraction is
established from the intensity of \ha\ emission from HVCs
with known distance bounds \citep{BM99,BM02,Pu03b}, assuming
that the \ha\ is photoionized.

Secondly, the disk spectrum in the range 912--2460\,\AA\ is derived from
the observed spectrum in the solar neighbourhood \citep{Me82}. Here
we make a key assumption that the observed field at the Sun's position
has already been extinguished with a mean 912--2460\,\AA\ opacity of 
$\tau$=1.5. This takes
account of the fact that some of the field will have originated close
to the O-B stars, and some from stars which have moved away from
the main star forming regions. We adopt the same opacity normal to
the disk such that 22\% of the soft UV photons escape into the halo.

The unextinguished 90-2460\,\AA\ ionizing spectrum in the disk is shown as the
uppermost green line in Figure 8. The extinguished spectra at distances of
10, 50, and 100\,kpc along the \pg\ sight line are then shown with the
lower curves.
%Note that there are 18 times more photons emerging in
%the soft field compared to the hard field. When the opacities are
%considered, this ratio increases to 67 along the polar axis in the far
%field limit (i.e., as $d\rightarrow\infty$). As the polar angle
%increases away from the polar axis, the halo 
%radiation field becomes systematically softer with the soft to hard UV
%photon flux ratio increasing to more than 100. This effect
%occurs because high extinction leads to isoflux contours that are more
%focused towards the polar axis \citep[see Appendix of][]{BH98}. 
The geometry of both the hard and soft MW radiation fields is illustrated in Fig. 9, where
the contours are lines of constant ionizing flux $\Phi$, and the
numbers given represent the logarithm of $\Phi$ in units of
photons\sqcm\,s$^{-1}$. 
The Galactic hard radiation field dominates the EGB hard radiation
inside the log\,$\Phi=4$ contour, which occurs at
approximately 40\,kpc for directions $|b|<30$\degr, 100\,kpc for
$30<|b|<60$\degr, and 180\,kpc for directions $|b|>60$\degr.
We do not including ionizing photons from hot white
dwarfs, planetary nebulae, and Galactic soft X-ray
sources, which may affect the MW radiation field
\citep{BH86}, or ionizing photons from the Magellanic Clouds.

%The disk-haloradiation field is axisymmetric around the $z$-axis\footnotemark, 
%\footnotetext{Including ionizing photons from the Magellanic Clouds
%  would add an asymmetry to the radiation field \citep[see][]{BM99}.}
%so only the Galactic latitude $b$ is
%needed to determine the ionizing flux as a function of distance. 
%The MW field emerges radially from the Galaxy so the one-sided
%approximation is valid. 

\subsection{General Properties of Photoionization models: Galactic vs
  Extragalactic} 

%To illustrate the different ionization patterns produced when our HVCs
%are subject to the two 
%radiation fields, we show 
%In Figure 10 we display the ionic column densities as a
%function of ionization parameter $U$ ($=n_{\gamma}/n_{\mathrm{H}}$,
%the ratio of the
%ionizing photon density to the gas density), for a cloud with a solar
%abundance pattern immersed in the EGB (left) and MW (right) fields. The MW
%ratios are valid for any distance from the Sun. The cloud is 
%chosen arbitrarily to have log\,$N$(\hi)=16.3 and
%[Z/H]=$-$0.50
%though the shape of the curves in insensitive to the these
%parameters, since each ionic curve scales equivalently for a solar
%relative abundance pattern, provided log\,$N$(\hi)$\lesssim$17.2.
%Note that the log\,$N$(H$_{tot}$) prediction on the plots has been shifted 
%downward by 3\,dex, for easier comparison with the other ions. 

%We note several important differences between the ionization patterns
%produced by the MW and those produced by the EGB, caused by differences
%in the detailed shape of the radiation fields. 
%The differences between the ionization patterns produced 
%by CLOUDY models compiled using 
%by the MW and EGB fields are shown
In Figure 10 we plot $N$(\ct)/$N$(\cw), $N$(\cf)/$N$(\cw),
$N$(\sit)/$N$(\siw), and $N$(\sif)/$N$(\siw) against $U$
($=n_{\gamma}/n_{\mathrm{H}}$) for three photoionization cases: pure MW,
pure EGB, and MW+EGB at 50\,kpc toward \he. 
%The doubly-ionized to 
%singly-ionized ratios (left panels) only
%differ between the MW and EGB fields at log\,$U>-2$ (low density), with
%the MW producing relatively higher ratios. The
%triply-ionized to singly-ionized ratios (right panels) show much more  
%difference between the two radiation fields; the MW field predicts
%lower ratios at log\,$U<-2$, but relatively higher ratios at
%log\,$U>-2$. 
The differences between these curves are caused by the EGB being
harder than the MW field. At log\,$U\lesssim-2.0$, this leads to higher
$N$(\sif)/$N$(\siw) and $N$(\cf)/$N$(\cw) ratios in the EGB. At
log\,$U\gtrsim-2$, the EGB begins to ionize silicon and carbon
beyond \sif\ and \cf, resulting in the turnover in the EGB curves
shown in Figure 10; the MW field produces
\sif\ and \cf\ as the dominant ionization stages at these ionization
parameters. 
%In all four panels the MW ratios at high
%log\,$U$ are higher than the EGB ratios; this effect arises
%since at log\,$U\gtrsim-2$, the EGB will ionize silicon and carbon
%beyond \sif\ and \cf, changing the fractions of the lower ionization
%stages in non-linear ways, whereas the MW produces
%\sif\ and \cf\ as the dominant ionization stages in this regime.
%to the carbon being ionized u
%to the \cw\ being mostly converted 
%into \ct\ at high log\,$U$, raising the $N$(\ct)/$N$(\cw) and
%$N$(\cf)/$N$(\cw) ratios; this turnover happens at higher log\,$U$ for
%the MW field. An identical argument holds for the silicon
%lines. for the EGB field, the turnover occurs at lower
%Since $N$(\cw)$\sim N$(\ct)$\sim N$(\cf) in typical 
%HVCs \citep[e.g.][]{Co04},
%Figure 10 can be used to determine the implied ionization parameter 
%for photoionized clouds, by measuring an ionic ratio and
%reading off the corresponding value of log\,$U$.
%Since we
%compare ionization stages from the same ion, these results
%are robust to changes in the relative elemental abundance pattern in
%the HVCs. 
The results are robust to increasing the number of
non-hydrogen-ionizing photons.
%, even though such photons are capable of
%ionizing \ion{C}{1} to \cw\ and \ion{Si}{1} to \siw. This is because
%almost all carbon and silicon atoms 
%in neutral interstellar clouds are in the singly ionized state anyway,
%so there is very little \ion{C}{1} and \ion{Si}{1} left to ionize.

%A separate difference between the EGB and MW models arises in the
%conversion from $U$ to $n_{\mathrm{H}}$; 
%Since the density of ionizing
%photons in the MW field at, for example, 10\,kpc is much higher than
%that of the EGB field, a given $U$ corresponds to a
%higher $n_{\mathrm{H}}$ in the MW case. 
%The $n_{\mathrm{H}}$ scales on the
%top of Figure 10 are therefore different in the two panels. 

\subsection{Modelling the HVCs toward \he\ and \pg}
%For modeling purposes we place the HVCs at distances of 10, 50, and
%100\,kpc in directions set by the Galactic coordinates or our sight
%lines, and scale the flux of the Galactic radiation using 
%Figure 9, to produce the lower green curves shown in Figure
%8. We ignore the attenuation of the EGB due to absorption by
%the \hi\ disk of the Galaxy. Table 6 shows the properties of the
%radiation fields at 
%these points along the sight lines to \he\ and \pg. 

We explicitly modeled each HVC absorption component toward \he\ and
\pg\ by comparing the CLOUDY model predictions
with the observed ionic column densities of carbon and silicon
(summarized in Table 5),
and determining the best-fit values of log\,$U$ and metallicity
[$Z$/H]\footnotemark. We assume all the \hi\ exists in the same phase as the low ions.
\footnotetext{[$Z$/H]=log\,($N_Z/N_{\mathrm{H}})-\mathrm{log}\,(N_Z/N_{\mathrm{H}})_\odot$.
  The relevant solar abundances we adopt throughout this section,
  using the definition 
  $A_{\mathrm{X}}$=$n(\mathrm{X})/n(\mathrm{H})$, are: 
  log\,$A_{\mathrm{C}}=-$3.61 \citep{AP02}; 
  log\,$A_{\mathrm{N}}=-$4.07 \citep{Ho01};
  log\,$A_{\mathrm{O}}=-$3.34 \citep{As04};
  log\,$A_{\mathrm{Si}}=-$4.46 \citep{Ho01};
  log\,$A_{\mathrm{Fe}}=-$4.55 \citep{Ho01}.}
This is repeated for assumed HVC distances of 10, 50, and
100\,kpc, with the Galactic radiation field scaled appropriately using
Figure 9. At each distance we add the uniform EGB contribution and enter the
combined field into CLOUDY (we do not account for the attenuation of EGB photons
passing through the \hi\ disk of the Milky Way). 
Table 6 shows the flux of the radiation field with distance along the sight lines
to \he\ and \pg.
%Since our best absorption line measurements are of lines of
%carbon and silicon, we study abundances using these two elements only.
%We began by trying to reproduce the column densities of all ionic
%stages observed, i.e. \cw, \ct, \cf, \siw, \sit, and \sif. $N$(\hi) is
%also provided as an input to the code in all cases.

\subsubsection{Low versus High Ions}
For all HVCs toward both \he\ and \pg, and for both EGB and MW radiation
fields, we found we could reproduce the column densities of the
singly- and doubly-ionized species (i.e., \cw, \ct, \siw, \sit), but
not the triply-ionized species (\cf\ and \sif) or the \os. In Figure 11 we show
the results from an example run, (HVC2 toward \pg), for two
different radiation field cases (EGB and MW at 10\,kpc).
The observations are shown as the larger data points, with 1$\sigma$
error bars. For either MW or EGB fiels, the observed levels of \sif,
\cf, and \os\ are 
significantly (at least 1\,dex) higher than the photoionization model
predicts at the value of $U$ 
which explains the singly and doubly ionized species. The
underproduction of the high ions is worse for the MW case. 
%For the EGB, the \cf\ prediction is typically 1.5 dex too low, and the
%\sif\ is short by approximately 1 dex; for the MW field, both the
%\cf\ and \sif\ prediction are typically two dex too low. \os\ is even
%more underproduced.
%: for all EGB photoionization models
%the \os\ prediction is around five dex lower than the observations; in
%the MW models, the underproduction of \os\ is even 
%more severe (since the MW flux sharply decreases above 54\,eV).
We attempted separate high-ion only models, but found that we could
not simultaneously reproduce the observed $N$(\sif), $N$(\cf), and
$N$(\os) with one value of $U$.
{\it These results strongly suggest that \sif, \cf, and certainly \os\
  are not photoionized in these HVCs}. The difficulty
in photoionizing \os\ has been reported elsewhere, both in the
modelling of observed HVC ionization patterns \citep[S03,][]{Co04,
  Co05, Ga05} and in theoretical work 
on the structure of photoionized clouds \citep{Ke99,GS04}. 
%Our results not only
%agree that \os\ in HVCs cannot be photoionized, but also suggest that \cf\ and
%some \sif\ cannot be either, at least in the same phase as the lower
%ions for the clouds under study.

Given that the species which appear most consistent with
photoionization are \cw, \ct, \siw, and \sit, we derive our
best-fit values of log\,$U$ and metallicity [$Z$/H] (assuming a solar
Si/C ratio) for each component by 
minimizing the $\chi^2$ statistic using data from these four ions
only. By investigating the depth of the $\chi^2$ minima, we find
typical errors on log\,$U$ and [$Z$/H] of $\pm0.2$\,dex (95\% c.l.). 
%To acknowledge the possibility of a non-solar Si/C ratio in the
%gas phase of these HVCs, we also investigate the effect of just
%fitting the carbon lines, or just the silicon lines. We therefore
%derive a  best fit [C/H] from \cw\ and \ct, a best fit [Si/H] from
%\siw\ and \sit, and 
%The last estimate provides a tighter constraint on $U$ since
%four data points are fit with one model. 
By combining [$Z$/H] and log\,$U$ with the
temperature and ionization breakdown for each ion returned by
the CLOUDY model, we derive the following quantities for the HVCs (see
Table 7):
gas density, physical size, thermal pressure, hydrogen
ionization fraction, and predicted intensity \ha\ emission. 

For a given component, the closer the clouds
are assumed to be to the Milky Way, the higher their density,
pressure, and predicted \ha\ intensity, and
the smaller their physical depth. 
The temperature of the photoionized
clouds remains confined  
to a range between 1.0--2.2$\times10^4$\,K.
% (for either radiation field
%for cases where log\,$U$=$-$5.0 to 0), so a higher gas density always
%corresponds to a higher thermal pressure.

\subsubsection{HVCs toward \he}
% Sort this section out!
For the four HVC components toward \he, we find overall metallicities of
$-0.6\pm0.2$, $-0.4\pm0.2$, $-0.9\pm0.3$, and $-0.6\pm0.2$; these
derived abundances are essentially independent of the assumed distance
to the clouds. 
Elemental abundance studies of the Magellanic Stream, lying eleven
degrees away from the \he\ sight line, have found
[Si/H]=$-$0.8$\pm$0.2 in the Leading Arm \citep{Se01b}, and
[\siw/H]$\gtrsim-1.5$ in the main body
of the stream \citep{Gi00}. Magellanic Stream abundances are similar to
those in the Small Magellanic Cloud\footnotemark, characterized by
%, [Si/H]$_{LMC}=+0.27$, [C/H]$_{LMC}=-0.35$
[Si/H]$_{SMC}=-0.51$, [C/H]$_{SMC}=-0.66$;
\citep[][corrected to our updated solar abundance
  scheme]{RD92}. 
%The Magellanic Stream is therefore closer to the
%SMC in metallicity\footnotemark.
\footnotetext{In the \citet{GN96} tidal model, all the gas in
the Magellanic Stream comes out of the SMC.}
Since our derived values for the silicon abundances
in the HVCs toward \he\ are suggestive of a Magellanic Stream origin, we are
able to place the HVCs at $\approx$50\,kpc.
Looking at the inferred densities, pressures, and  
cloud sizes of the MW+EGB 50\,kpc models of the \he\ clouds in Table
7, we find that 
the four HVCs have derived densities in the range
log\,$n_{\mathrm{H}}$=$-$2.3 to $-$2.1, physical sizes of 39 to 120\,pc,
thermal pressures of
95--190\,cm$^{-3}$\,K\footnotemark, and hydrogen ionization fractions of 98\%.
If we assume spherical symmetry for the clouds, they
have masses in the range 10-250\,$M_{\odot}$ and angular sizes on the
sky of several arcminutes. % Checked OK! 
Other geometries could result in substantially larger values for both the
mass and angular size.
\footnotetext{The main body of the Stream contains
  clumps with higher thermal pressure: \citet{Se01b} measure 
  $P/k\sim360$\,cm$^{-3}$\,K in the leading-arm.}
%This similarity in their
%physical conditions suggests that these clouds share a related
%origin. 
One explanation is that the clouds trace material
tidally stripped from the main (21-cm) body of the Magellanic Stream,
which are then ionized by the pervading MW radiation field.
 %, this implies that the Stream does not
%have a well-defined edge, but is instead surrounded by a flocculent
%population of small clouds. %We
%note that \citet{Ho04} have used sensitive Arecibo observations to
%detect ``mini-HVCs'' in \hi\ emission -- we may be detecting the same
%population of clouds in absorption.

If we treat the detection of \few\ in HVC2 at 2.4$\sigma$
significance as real, then the CLOUDY model returns [Fe/H]$\approx$
+0.6 in this component, whereas carbon and silicon give [Z/H] =
$-$0.4. We note that 
\citet{Lu98} concluded that dust was present in the Magellanic Stream, 
since the measured S/Fe ratio of eight times the solar value in
HVC~287.5+22.5+240 requires much of the iron to be depleted out of the
gas phase. Unless the elemental abundances are markedly different in
the core toward HVC~287.5+22.5+240 than in the HVCs under study here,
we conclude that the iron feature in HVC2 is likely not real; it could
be due to fixed pattern noise.

\citet{Co05} report a size of 5.9\,kpc for the same HVC system toward
\he\ that we model here, based on different photoionization models. 
The difference between their result and ours (HVC sizes of 39--120\,pc if the
clouds are at 50\,kpc) is partly due to their modeling the system
as a single component, whereas we use four components, with smaller
individual column densities. A second reason is metallicity: we derive 
values higher than
the 0.1 solar assumed by \citet{Co05}, resulting in our finding lower ionization
parameters, higher densities, and smaller cloud sizes.
%different assumptions about component structure (single cloud vs four
%clouds) and different assumptions about metallicity (0.1 solar vs free
%parameter), and also different radiation fields (power law vs Haardt \& Madau).

\subsubsection{HVCs toward \pg}
The metallicity of the 125\kms\ HVC toward \pg\ is
[$Z$/H]=$-$0.8$\pm$0.2 independent of distance, suggesting it is not a
fountain cloud,
which would be expected to have solar metallicity or above, but rather
a cloud of non-Galactic origin that happens to be moving away from the
LSR. Our modeling finds the cloud to be 99--100\% ionized.
The detection of \oi\ at apparent 3.0$\sigma$ significance in HVC2 
implies [\oi/\hi]$\approx$[O/H]=+0.6$\pm$0.2 if the feature is real. Since the
carbon and silicon lines suggest metallicities lower by $\approx$1.6\,dex, this
\oi\ detection is likely spurious. We have no information on the
distance to this absorber.
%, but our models show the photoionized gas to
%be 99\% ionized whatever the distances.
%, so it is difficult to constrain the
%physical conditions in the cloud.

%The $-$150\kms\ HVC also has a low metallicity
%([$Z$/H]=$-$0.5$\pm$0.2). 
The $-$146\kms\ HVC has a metallicity of $-$0.6$\pm$0.2.
The \siw/\sit\ and \siw/\cw\ column density
ratios (and overall line strengths) in the $-$150\kms\ HVC
toward \pg\ are similar to those of the +100\kms\ HVC in the
spectrum of PG~1116+215 recently presented by \citet{Ga05}.
The properties of both of these HVCs, including these low-ion ratios,
the absence of \cf\ and \os, and the approximate \hi\ columns,
are highly reminiscent of the Virgo Cluster
Ly$\alpha$ cloud in the spectrum of 3C~273.0 at $z_{abs}$ = 0.00530
\citep{Se01a, Tr02}. Tripp et al. find that the
3C~273.0 Ly$\alpha$ cloud is quite small, $\sim$70\,pc, of the same order
as the sizes we derive for the $-$146\kms\ HVC (80\,pc if at 100\,kpc).
%, suggesting it may be an HVC itself. 
\citet{Tu05} report two Ly$\alpha$ absorbers toward QSO PG~1211+143,
each lying
within 150\,kpc of a potential host galaxy, which also have ionization
properties similar to those of Galactic highly ionized
HVCs. 
%The Virgo Ly$\alpha$ absorber may be an HVC like the \pg\ and
%PG~1116+215 HVCs. 
%These clouds would require pressure confinement
%by an ambient medium.

\subsection{HVC Thermal Pressure}
We compare our derived HVC pressures with the estimated pressure of an
isothermal Galactic corona, as a function of distance. The coronal pressure has
been calculated by \citet{St02} for a $2\times10^6$\,K
hydrostatic halo using a two-component (disk and dark-matter) Galactic
potential. Sternberg et al. find
$P_{HIM}/k\approx$[1000, 250, 100]\,cm$^{-3}$\,K at
$r=$[10, 50, 100]\,kpc. By averaging the results of the six HVCs we
model, we find a typical $P/k$ of [530, 140, 50]\,cm$^{-3}$\,K if the
clouds are at these 
same distances. Ignoring the difference between the Sun-HVC and
Galactic Center-HVC distances, our clouds would therefore be 
slightly underpressurized in a $2\times10^6$\,K corona, but would be
close to thermal pressure equilibrium in a $1\times10^6$\,K corona.
Such a medium is suspected from UV \citep[S03,][]{Fo04, Co04,
  Co05, Ga05} and X-ray high-ion
absorption \citep{Fa03, Mc04} studies, and from pulsar dispersion measures,
X-ray emission, and properties of the Magellanic Stream \citep[see][and
  references therein]{QM01}.
At 50\,kpc, the density of a fully ionized $10^6$\,K corona would be
$n_{\mathrm{H}}=4-9\times10^{-5}$\,cm$^{-3}$ if thermal pressure
equilibrium holds in the HVCs toward \he, assumed to be part of the
Magellanic Stream. Magnetic fields and cosmic rays may also provide
pressure confinement, but the similarity of thermal pressures between
a hot corona and these HVCs is striking nonetheless.

\section{Collisional Ionization: the Cloud Boundaries}

%In \S6 we established that photoionization can explain the observed
%column densities of singly and doubly ionized species. However,
%regardless of whether the clouds are nearby and subject to escaping
%Milky Way radiation or exposed to the EGB, photoionization cannot account
%for the observed high-ion column densities. 
Since photoionization cannot explain the high ions, we now investigate
collisional ionization models, specifically: collisional ionization
equilibrium \citep[CIE;][]{SD93}, magnetized conductive 
interfaces \citep[CIs;][]{Bo90}, turbulent mixing layers \citep[TMLs;][]{Sl93},
radiatively cooling hot gas \citep[RC;][]{EC86}, and shock ionization
\citep[SI;][]{DS96}. A brief description of each of these
models, and the appropriate range of their parameter space, is given in
\citet{Fo04}. 
%\footnotetext{In the RC model high stages of ionization are
%  formed from the recombination of even higher stages of ionization
%  existing in hot gas; however we include it here since
%  that hot gas was collisional ionized in the first place.} 

To increase the sample of observations over the high-ion
detections in HVCs toward \he\ and \pg, we produce a compilation of all
measurements of highly ionized HVCs 
%components\footnotemark
%\footnotetext{We use the word ``component'' to distinguish discrete
%  \os\ absorption line features from \pvws.}
where \sif, \cf, and \nf\ data exist \citep{Fo04, Co04, Ga05}. This
compilation is presented in Table 8. In cases where only a limit could be
measured for $N$(\sif), $N$(\cf) or $N$(\nf), we present a corresponding 
3$\sigma$ upper limit to the ratio of that ion to \os. We calculate
the mean and standard deviation of the high 
ion ratios in these fourteen HVCs, and compare these to the integrated
averages measured in the Galactic halo \citep{Zs03}. 

There is an enormous range in $N$(\hi)/$N$(\os) observed in HVCs, from
$\sim\!10^6$ in Complex C to $\sim\!10^2$ in components toward \pg\ and
PKS~2155-304. The variation in the ratio is primarily due to variations in
$N$(\hi), whereas $N$(\os) is always found to be within 1\,dex of 13.0. This
represents a crucial piece of evidence in understanding the structure
of HVCs: no matter how much neutral gas there is, some process
regulates the production of highly ionized gas and always produces a
similar amount of \os. A corresponding dispersion
in the $N$(\hi)/$N$(\os) ratio in low redshift intergalactic \os\
absorbers has been noted \citep{Tr00, DS05}.
%; the latter authors use the
%term ``multi-phase ratio'' to describe this effect.  

% Could cut this paragraph
%The sample of high ion-high ion ratios show much less dispersion. The
%$N$(\cf)/$N$(\os) and $N$(\sif)/$N$(\os) HVC averages are not significantly
%different from the Galactic halo sight line averages, suggesting that
%similar ionization mechanisms may be at work in HVCs as in the thick
%disk of the Milky Way's ISM. There is a somewhat larger dispersion in
%$N$(\cf)/$N$(\sif) seen in our small sample of highly ionized HVCs
%than in the halo: HVC values range from 1.1 to $>\!25$, whereas the
%integrated halo values lie between $2.0\pm1.0$ and $4.7\pm0.9$
%\citep{Zs03}. However, we note that high-resolution
%observations have found absorption components in the Galactic halo where the
%$N$(\cf)/$N$(\sif) ratio is significantly higher \citep[$5.1\pm0.6$ toward
%HD~119608;][]{Se97}, and also where the ratio is much lower
%\citep[$0.7\pm0.4$ in a narrow component toward HD~116852;]{Fo03}, than the
%halo average, so it is possible that the true dispersion in ionic ratios may be
%as large in the Galactic halo as in HVCs, reflecting multiple
%ionization processes in both locations. 

Our observations are compared with the model predictions in Figure 12, which
contains plots of $N$(\cf)/$N$(\os) vs  
$N$(\nf)/$N$(\os) (left panel), and $N$(\sif)/$N$(\os) vs
$N$(\nf)/$N$(\os) (right panel). All measurements of high-ion column density
ratios in HVC components (given in Table 8) are shown in red, upper
limits are shown in purple, and the 
Galactic halo average is shown in green \citep{Zs03}. The shaded regions
represent the range of parameter space allowed by the various
models. The blue arrows show the effect 
of correcting the model predictions to a scheme of SMC abundances
\citep{RD92}, assuming that oxygen is the dominant coolant
\citep[see][for a description of the correction process]{Fo04}. A
summary of the ability of 
each model to reproduce the observations is summarized in Table 9.
%shock ionization and conductive interfaces are most successful.
%, where we present the statistics of what fraction of observations can
%be explained by each model.

%Neither radiative cooling of a hot gas phase \citep[as modelled
%by][]{EC86}, or CIE with solar abundances can explain any of the
%data points on Figure 13, nor can they explain the kinematic
%alignment of highly ionized and neutral gas. Turbulent mixing layers
%\citep[as modelled by][]{Sl93} can explain the $N$(\sif)/$N$(\os) and 
%$N$(\cf)/$N$(\sif) ratios but overestimate $N$(\cf)/$N$(\os) in 5 out
%of 10 cases. 

The shock ionization model, when calculated using a shock
speed of 200\kms\ (closest to the LSR velocities of the HVCs
toward \he\ and \pg) systematically underestimates
$N$(\cf)/$N$(\os). A much higher velocity
shock ($\sim500$\kms) is needed to explain the $N$(\cf)/$N$(\os)
ratio by shock ionization, which makes this scenario unlikely, though we
cannot rule it out, since a HVC falling in at 200\kms\ against a
300\kms\ Galactic outflow would produce a relative velocity of
$\sim500$\kms, and hence a higher Mach number. The region of
ratio-ratio space in Figure 12 covered by the shock model prediction
allows for the possibility of shock velocities of up to 500\kms.

%\subsection{Conductive Interfaces}
Conductive interfaces can explain the $N$(\cf)/$N$(\os) and
$N$(\nf)/$N$(\os) ratios in almost all cases, but cannot explain the
$N$(\sif)/$N$(\os) observations. However, less weight should
be given to the ratios involving $N$(\sif).
%, an ion that can be
%photoionized from \sit\ by a photon having $E>33.5$\,eV. 
Ionizing photons produced in the interface itself \citep{SS92, Sl04}
might inflate $N$(\sif), rendering the purely collisional ionization model
predictions invalid. In particular, 
the interface should be a strong emitter of \ion{He}{2} $\lambda304$
radiation, capable of ionizing \sif\ but not \cf\ \citep{SS92, Sa94}. 
Such ionizing photons are {\it not} accounted for in our
CLOUDY photoionization models of \S6. In the case where all ratios
involving $N$(\sif) 
are ignored, the conductive interface model is the most successful at
explaining the remaining ratios (\cf/\os\ and \nf/\os) in HVCs (and
the average Galactic halo value). The interface model is also consistent with
the observed HVC absorption kinematics, since the observed high
ion-low ion offsets are of the same order (0--25\kms) as would be
expected in an evaporative flow off a conductive interface \citep{BH87}.
To further our understanding of conductive
interfaces, the effects of HVC motion on the
conduction across the interface need to be explored
theoretically. Numerical simulations of the interaction between
infalling HVCs and surrounding media \citep{QM01, ML04} indicate that
cometary features, turbulence, and bow shocks are
all expected to occur around HVCs. 
%Whether conduction is still
%efficient in these scenarios is unknown, but beyond the scope of this
%paper. 
More advanced models of conductive interfaces would also allow 
for a full treatment of the ionizing radiation produced 
within the interface itself \citep[the diffuse field;][]{FF94, Sl04}. 
Such a treatment would be able to make predictions on the relationship
between HVC \os\ absorption and HVC \ha\ emission \citep{WW96, Tu98,
Tu02, Pu03b}.
%, which has been interpreted as arising in photoionized
%boundary layers.
If a significant fraction of the \ha\ emission from HVCs arises from
conductive interfaces, rather than by photoionization from escaping
Galactic radiation, then the true escape fraction of Galactic
ionizing photons may be less than the 6\% (normal to disk) discussed
in \S6. Further \ha\ studies of HVCs will help elucidate this issue.
%\ha\ emission line studies of HVCs can also be used to verify the
%escape fraction of the MW radiation field presented in \S6.
%(although shock ionization cannot currently be 
%ruled out). 
%Improved models of the radiation field existing within HVC
%interfaces could then be tested with simultaneous optical/UV studies of
%\ha-emitting and \os-absorbing HVCs. 
%Finally, a careful analysis of 
%what percentage of discrete \hvo\ absorbers are accompanied by high-velocity
%\hi\ absorbers will reveal how successful the interface scenario can
%be at explaining highly ionized HVCs.

\section{Positive-Velocity Wings}
The \os\ profile toward \pg, with its smooth \pvw\ rather than discrete
component, is qualitatively different from the 
\cw, \ct, \cf, \sit, and \sif\ absorption line
profiles in this direction.
%, which are more component-like. 
%Is there a
%separate physical process generating the \os\ wing?
%We consider two possibilities: (i) the \os\
%wing represents a blend of unresolved components, and (ii) the \os\
%wing traces a hot, highly ionized outflow. 
%One possible way to produce the \os\ apparent column density profile
%in Fig. 7 is with two or more blended components. 
The \cf\ and \sif\ HVCs, although observed at low significance, each appear to show
a broad component centered near 150\kms. One interpretation of the
data is that the 125\kms\ HVC (seen in \cw) is surrounded by highly ionized
interface layers, which contain the ions \cf\ and \sif. This
explanation is consistent with the \cf/\sif\ ratio, the \cf-\cw\
offset, and our explanation for the high ions in the \he\ clouds
and other HVCs, but poses the question: where is the \os\ in this interface?
%the interface \os\ blends with
%the rest of the \hvo\ absorption to produce the appearance of a
%``wing''. 
%The observed \os\ profile is consistent 
%with this idea, 
We show using the blue line in the upper-right
panel of Figure 7 that the interface \os\ could well be hiding
underneath the wing. This line shows the contribution to the \os\
profile of an interface component having the same width as the \cf\ HVC
and a typical $N$(\cf)/$N$(\os) interface ratio (of 1). As can be seen, such a
component can explain all the \os\ absorption between 140 and
200\kms, but {\it not the remaining \hvo\ between 75 and 140\kms.}
%, but in this interpretation, the idea of the ``wing'' as a
%single macroscopic flow is discarded. 
%The 25\kms\ offset between the
%\cw\ and \cf\ absorbers may trace an evaporative flow off the face of
%the cloud; offsets of this kind have been seen toward PG~1116+215
%by \citet{Ga05}.%, who explain them as the result of a cloud moving
%through a hotter medium. 
%if they trace the cloud 
%interface and cloud core, respectively.
%If evaporative flows off the cloud are the explanation for the 25\kms\
%offsets, then there should be flows off the
%front and rear faces of the cloud, giving two components, but only one
%component is seen in the data.

%The \pvws\ observed toward \pg, 3C~273 \citep{Se01a}, Mrk~421
%\citep{Sa05a}, and other targets (S03) 
The \os\ seen at 75--140\kms\ toward \pg\ has no accompanying absorption
seen in \cf\ or \sif: we find
log\,$N$(\os)=13.67$\pm$0.08, log\,$N$(\cf)$<$13.22, and
log\,$N$(\sif)$<$12.49 over the same velocity range. Assuming solar
C/O and Si/O ratios and CIE 
conditions apply, the \cf/\os\ and \sif/\os\ ratios imply log\,$T>5.31$ and $>5.26$,
respectively. This hot gas may be tracing a hot Galactic Fountain 
\citep{SF76, Br80, NI89, HB90}: a
circulation of interstellar material in which supernovae-heated gas
rises buoyantly into the halo, cools, and then rains back toward the
disk. In a hot fountain, little \os\ is
expected in the initial 10$^6$\,K flow, but as the gas
cools, higher stages of ionization will recombine into O$^{+5}$ as the
flow drives outward, even though gas at $3\times10^5$\,K (where \os\
is formed by collisions) is not hot enough to drive the flow. The
model is attractive for explaining \pvws\ since it naturally explains why the
wings are always seen to be redshifted; no negative-velocity (infalling)
wings are seen in \os\ profiles. 
%The model also explains the smooth
%shape of \pvw\ absorption profiles, which decrease 
%smoothly with increasing velocity. 
%A random series of interface % components would produce
%a flat \os\ profile. 
%This distinct kinematic signature
%of all 22 \pvws\ in the S03 survey suggests they are more
%likely produced by a flow rather than a sum over many simple components.

The Galactic Fountain model has had some success explaining the
kinematics and metallicites of the intermediate velocity clouds
(IVCs), defined to be at $|v_{LSR}|=30-90$\kms\ \citep{Ri01}. One could
speculate that hot outflowing fountain gas is detected as \pvws, and
cooler inflowing gas is 
detected in the form of neutral IVCs. In this case, the outflowing gas
must be dense enough to cool in the time it takes to rise out into the
halo, otherwise the \hi\ IVCs will never condense. Since the instantaneous
radiative cooling time is 
$t_{cool}=2.4\times10^8n^{-1}_{-3}T_6\Lambda^{-1}_{-21.9}$\,yr
\citep{HB90}, where $\Lambda_{-21.9}$ is the cooling function in
units of its value at $10^6$\,K ($1.26\times10^{-22}$\,erg\,cm$^3$\,s$^{-1}$),
$T_6\equiv T/10^6$\,K, and $n_{-3}\equiv n/10^{-3}$ cm$^{-3}$, we find
that a $10^6$\,K fountain flow moving at 100\kms\ must
have a density $n_{-3}\approx 9$ to radiate half its internal energy
in a sound crossing time (28\,Myr for a scale height
of 2.5\,kpc, appropriate for an isothermal $10^6$\,K halo), which
approximates the  time taken for the gas to fill the halo. 
%If $n_{-3}\gg9$, the gas
%would cool very quickly and we would not expect to see \os\ wings.

A related possibility is that the wings
trace an escaping Galactic wind, energetic enough
%, a high velocity component of
%upwelling material with enough kinetic energy 
to escape the Galactic potential well. Cosmic rays could play a role
in driving such a Galactic wind and supporting
high-$z$ gas \citep{Br91}. 
%\footnotemark. 
%\footnotetext{\citet{Ca88} find $v_{esc}\approx500$\kms\ in the
%  Galactic plane, decreasing with $z$.} 
\os-containing outflows are predicted by
hydrodynamical simulations of galaxy winds \citep{MF99, SS00}, and
have been detected in starburst galaxies
NGC~1705 \citep{He01} and NGC~625 \citep{Ca04}.
%(albeit in the form of discrete components, not wings).
Mid-infrared images of the Galactic
center show a large-scale bipolar outflow in our own galaxy \citep{BC03}. 
Analysis of a larger sample of \pvws\ will be necessary to fully
explore the properties of an \os\ outflow.

%The behavior of the \os\ wing toward \pg\ can be compared
%with the wings investigated by \citet{Sa05a} for the nearby sight line
%toward Mrk~421 ($l=180\degr, b=+65\degr$). These authors found a very
%close kinematic coupling between \ct\ and \os, with
%$N$(\ct)/$N$(\os) ratio constant at $\approx$0.1 over the velocity
%range of the wing; we also find a reasonably 
%constant $N$(\ct)/$N$(\os) ratio in the \pg\ wing, but with a much
%higher value of 1.0, implying a lower temperature than the Mrk~421
%wing. 
%Further observations investigating the distribution of \pvws\ over the sky are
%necessary to investigate the patchiness and physical conditions of a
%hot Galactic outflow.
%The sight line to QSO 3C~273.0 ($l=290\degr, 
%b=+64\degr$) is another Northern direction where a \pvw\ has
%been detected in \os\ absorption \citep{Se01a}. 

%There is no evidence for
%corresponding absorption in \hi\ or any metal lines in this direction.
%although the presence of \ct\
%cannot be assessed because of a blend with a redshifted IGM line.
%Sembach et al. conclude that a Galactic outflow is the most likely
%explanation for the wing toward 3C~273, given that strong
%X-ray emission in this direction suggests that energetic activity may
%be venting hot gas into the halo.

%In an upcoming paper, we will explore the properties of the full
%sample of \pvws\ detected in the S03 survey. Since some complete sight
%lines through the halo do not have wings (S03), and neither do many halo
%stars up to distances of several kpc \citep{Zs03}, this further study
%will investigate the patchiness of any such outflow.

\section{Summary}
We have studied \fuse\ and \hst/STIS spectra to investigate the
ionization, kinematics, elemental abundances, and physical conditions in the HVCs
toward \he\ and \pg. We summarize our study in the following key points. 

\begin{enumerate}

\item {\bf HVCs toward \he.} HVC absorption in the range 80--230\kms\
  is detected in the \he\ direction in lines of \hi, \cw, \ct, \cf,
  \os, \siw, \sit, and \sif. Four components are clearly seen in
  this velocity range in the high-resolution STIS observations of \cw, \siw,
  and \sit, at 99, 148, 175, and 193\kms. Each of these components is
  accompanied by a broader, highly ionized 
  (\cf\ or \os) component, indicating that warm and hot phases exist in each
  component.
  %, since the high- and low-ionization gas phases are not
  %co-spatial. 
  The warm high-velocity clouds are all found to have
  similar densities, pressures, physical sizes, and ionization levels;
  their metallicities  
  ([Z/H] between $-$0.9 and $-$0.4) indicate the clouds may be outlying
  fragments of the Magellanic Stream, whose main body lies 11\degr\
  away on the sky at similar velocities to the \he\ absorbers. 

\item {\bf HVCs toward \pg.} HVCs exist at both negative and positive
  velocities toward \pg. The negative-velocity HVC at $-$146\kms\ is seen in \cw,
  \siw, \sit, and \sif, has [Z/H]=$-0.6\pm0.2$, and contains no detectable
  \cf\ or \os. At positive velocities,
  a narrow discrete HVC component with [Z/H]=$-0.8\pm0.2$ is seen in \cw\ at
  124\kms. \ct\ and 
  \sit\ show much broader absorption components centered at the same
  velocity. The highly ionized species \sif\ and \cf\ show 
  a weak component near to 150\kms, offset from the lower ionization species by
  $\sim25$\kms. The \os\ profile contains a smooth \pvw\
  extending from 75--200\kms. We show that part of this \os\ absorption
  could be produced in the same absorber that contains \cf\ and
  \sif, but the gas in the range 75--140\kms\ requires an alternative explanation. 
  %with no evidence for the discrete 125\kms\
  %component. The component and wing are likely unrelated phenomena:
  %the component has a low metallicity ([$Z$/H]$=-1.0\pm0.2$) so it is likely
  %extragalactic material. An interface around
  %the component could explain the \cf\ and \sif\ profiles,
  %if the corresponding interface \os\ component is hidden underneath the
  %wing -- we show this is consistent with observations. %; its
  %ionization properties are similar to those of 
  %Ly$\alpha$ clouds in the Virgo cluster.

\item {\bf CLOUDY Photoionization Modeling.} 
  %The column densities of
  %the \hi\ components of HVCs toward \he\ and \pg\ are in the range
  %$10^{16}$--$10^{17}$\sqcm, indicating the clouds are optically thin to
  %Lyman continuum radiation. 
  %We model the HVCs as subject to either the
  %extragalactic background or the escaping Galactic
  %ionizing radiation, derived from the spiral distribution of O-B
  %stars, measurements of the soft stellar radiation field, and an
  %estimate of the photon escape fraction. 
  The Galactic ionizing radiation
  field from hot stars dominates the EGB ionizing radiation inside the
  log\,$\Phi_{<912}=4$ contour, which occurs at
  approximately 40\,kpc for directions $|b|<30$\degr, 100\,kpc for
  $30<|b|<60$\degr, and 180\,kpc for directions $|b|>60$\degr.
  Photoionization is a good explanation for the \cw, \ct, \siw, and
  \sit\ in the HVCs.
  Neither the EGB or the MW field can account for \cf\ and
  \os\ in the HVCs, confirming that they exist in a separate, hotter
  gas phase. 
  %We present new calculations of
  %the silicon and carbon ionization balance in clouds ionized by
  %escaping Milky Way photons. 

\item {\bf Highly Ionized HVC Components.} 
  %Since photoionization
  %cannot account for the high ions, we compare the high-ion
  %observations with the predictions of both equilibrium and
  %non-equilibrium collisional ionization models. \
  After studying the
  ionic column density ratio and kinematic information of thirteen
  highly ionized HVC components reported here and in the literature,
  we believe the
  highly ionized HVC components arise at the interfaces between the 
  cool/warm HVCs and a surrounding medium. Conductive interfaces
  can reproduce $N$(\cf)/$N$(\os) and $N$(\nf)/$N$(\os) in 11/12
  highly ionized HVCs, though there are discrepancies remaining to be
  resolved with the $N$(\sif) prediction. HVC high-ion column density
  ratios are not significantly different from the Galactic Halo
  averages of \citet{Zs03}, suggesting that similar ionization
  processes are at work in both environments. The interface scenario
  implies the existence of a hot, extended corona to the Milky Way 
  extending out 50\,kpc to the Magellanic Clouds.
  %, as suggested by
  %several authors \citep[S03,][]{Fo04,Ga05,Co05}. 
  Such a corona can also provide
  confinement to the warm HVCs; our modelling shows the \he\ HVCs would be close to pressure equilibrium in
  a fully ionized $10^6$\,K corona with $n_{\mathrm{H}}=4-9\times10^{-5}$\,cm$^{-3}$ at
  50\,kpc.
 
\item {\bf Positive-Velocity Wings.} The extended \os\ absorption wing seen
  between 75 and 140\kms\ toward \pg\ may trace a hot Galactic
  outflow, because of the asymmetric profile and the lack of
  accompanying wing absorption in less-ionized species. The lack of
  \cf\ and \sif\ in this velocity range can be used to constrain the
  temperature in the wing gas to be log\,$T>5.31$, assuming CIE and solar
  elemental abundance ratios.

\end{enumerate}

The authors acknowledge an anonymous referee for a perceptive
report. We thank Marilyn Meade for running the CALFUSE
pipeline. The STIS observations of \he\ were obtained for HST program
9148, with financial support through NASA grant HST-GO-9184.08-A from
the Space 
Telescope Science Institute. BPW was supported by NASA grants
NNG04GD85G and NAG5-7444. BDS acknowledges support from NASA grant
NNG04GC70G. TMT appreciates additional support from NASA
LTSA grant NNG04GG73G.

\begin{deluxetable}{lcccc ccc}
\tablewidth{0pt}
\tabletypesize{\footnotesize}
\tablecaption{Sight Line Properties}
\tablehead{Target & Type & $z$ & $l$ & $b$ & $V$ & $F_{1030}$\tm{a} &
  $\Delta v_{LSR}$\tm{b}\\
                &      &     &(\degr)&(\degr)& (mag.) & (flux units) & (\kms)}
\startdata
\he  & Sey1 & 0.495 & 253.94 & $-$65.78 & 15.2 & 2.7 & $-14.3$\\
\pg  & Sey1 & 0.234 & 179.79 & +51.71   & 14.5 & 5.1 & $-0.3\phn$\\
\enddata
\tablecomments{Information retreived from the NED database
  (http://nedwww.ipac.caltech.edu), unless noted below.} 
\tn{a}{Measured flux at 1030\,\AA; 1 flux unit =
  $10^{-14}$\,erg\sqcm\,s$^{-1}$\,\AA$^{-1}$.} 
\tn{b}{Correction from heliocentric to LSR velocity frame. $\Delta
  v_{LSR}=v_{LSR}-v_{Helio}$.}
\end{deluxetable} 

\begin{deluxetable}{lcclc r}
\tablewidth{0pt}
\tabletypesize{\footnotesize}
\tablecaption{Summary of Observations}
\tablehead{Instrument & Target & Aperture & Dataset ID & Observation Date &
  $t_{exp}$\\ 
 & & & & & (ks)}
\startdata
\hst/STIS\tm{a} & 
\he & $0.2\arcsec\!\times\!0.06\arcsec$ &  
   O6E107020,30    & 2002 Dec 26 & 6.0\\
&&&O6E108010,20,30 & 2002 Dec 26 & 8.2\\
&&&O6E109020,30    & 2002 Dec 27 & 6.0\\
&&&O6E110010,20,30 & 2002 Dec 29 & 8.2\\
&&&O6E111010,20    & 2002 Dec 31 & 5.1\\
&&&O6E111030,40    & 2003 Jan 01 & 6.0\\
&&&&&                    Total=39.6\\
& \pg & $0.2\arcsec\!\times\!0.2\arcsec$\phn  & 
   O4X001010,20 & 1998 Dec 11 & 13.7\\
&&&O4X002010    & 1998 Dec 04 & 10.7\\
&&&&&         Total=24.4\\
\tableline
\fuse\tm{b} &
\he & $30\arcsec\!\times\!30\arcsec$ & 
   P2071301 (v2.1.6) & 2000 Dec 12 & 11.0\\
&&&P1019101 (v2.1.6) & 2001 Oct 03 & 50.3\\
&&&P1019102 (v2.4.0) & 2002 Nov 15 & 14.5\\
&&&P1019103 (v2.4.0) & 2002 Nov 16 & 18.9\\
&&&P1019104 (v2.4.0) & 2002 Nov 17 & 18.1\\
&&&D0270101 (v2.4.0) & 2003 Sep 01 & 23.9\\
&&&D0270102 (v2.4.0) & 2003 Sep 03 & 39.9\\
&&&D0270103 (v2.4.0) & 2003 Oct 31 & 16.7\\
&&&&&  Total=193.3\\
& \pg & $30\arcsec\!\times\!30\arcsec$ & 
   P1012202 (v2.4.0) & 1999 Dec 30 & 38.9\\
&&&P1012201 (v2.1.6) & 2000 May 04 & 36.5\\
&&&&&           Total=75.4
\enddata
\tn{a}{All STIS observations were taken with the E140M grating,
  providing 6.7\kms\ resolution (FWHM). The unbinned pixel size is
  between 3.0 and 3.2\kms (varying with wavelength).}
\tn{b}{The \fuse\ resolution is $\approx20$\kms\ (FWHM) in the LiF
  channels, and $\approx25$\kms\ in the SiC channels. Unbinned pixels 
  are 2.0\kms\ wide. We list the CALFUSE pipeline version used in parentheses after
  the Dataset ID. The difference between v2.1.6 and v2.4.0 is negligible.} 
\end{deluxetable}

\begin{turnpage}
\begin{deluxetable}{lcccc ccccc ccccc ccc}
\tablewidth{0pt}
%\rotate
\tabcolsep=2pt
\tabletypesize{\tiny}
\tablecaption{Measurements of HVC Absorption toward \he}
\tablehead{Line & S/N\tm{a} 
& \multicolumn{4}{c}{\underline{\phm{aaa}HV1,  80 to 125\kms}\phm{aaa}}
& \multicolumn{4}{c}{\underline{\phm{aaa}HV2, 125 to 160\kms}\phm{aaa}}
& \multicolumn{4}{c}{\underline{\phm{aaa}HV3, 160 to 185\kms}\phm{aaa}}
& \multicolumn{4}{c}{\underline{\phm{aaa}HV4, 185 to 230\kms}\phm{aaa}}\\
&
& $\bar{v}$\tm{b} & $b$\tm{c} & $W_{\lambda}$\tm{d} & \phm{aaa}log\,$N_a$\tm{e}
& $\bar{v}$\tm{b} & $b$\tm{c} & $W_{\lambda}$\tm{d} & \phm{aaa}log\,$N_a$\tm{e}
& $\bar{v}$\tm{b} & $b$\tm{c} & $W_{\lambda}$\tm{d} & \phm{aaa}log\,$N_a$\tm{e}
& $\bar{v}$\tm{b} & $b$\tm{c} & $W_{\lambda}$\tm{d} & \phm{aaa}log\,$N_a$\tm{e}}
%(km\,s$^{-1}$) & (km\,s$^{-1}$) & (m\AA) & ($N_a$ in \sqcm)}

\startdata
\hi $\,\lambda$920.963 & 10 & \phn92$\pm$5 & 25$\pm$2 & 136$\pm$07$\pm$08\phn & 16.29$^{+0.02+0.04}_{-0.02-0.04}$&
                           147$\pm$6 & 14$\pm$2 &  88$\pm$05$\pm$10 & 16.16$^{+0.06+0.18}_{-0.07-0.31}$& 
                           172$\pm$5 & 12$\pm$1 &  85$\pm$03$\pm$17 & 16.20$^{+0.08+0.30}_{-0.11-0.30}$&
                           201$\pm$6 & 19$\pm$3 & 107$\pm$07$\pm$19 & 16.25$^{+0.04+0.19}_{-0.05-0.36}$\\

\hi $\,\lambda$919.351 & 9 & \nodata & \nodata & \tm{f} & \tm{f} & 
%95$\pm$6 & 24$\pm$3 & 114$\pm$08$\pm$05 & 16.29$^{+0.05+0.02}_{-0.06-0.02}$&
                           146$\pm$7 & 15$\pm$2 &  79$\pm$05$\pm$06 & 16.21$^{+0.05+0.05}_{-0.07-0.06}$&
                           173$\pm$6 & 12$\pm$1 &  72$\pm$04$\pm$11 & 16.26$^{+0.07+0.07}_{-0.07-0.08}$&
                           203$\pm$6 & 19$\pm$2 & 110$\pm$07$\pm$14 & 16.39$^{+0.03+0.06}_{-0.04-0.07}$\\

\hi $\,\lambda$918.129 & 8 & \phn90$\pm$7 & 27$\pm$2 &  97$\pm$12$\pm$07 & 16.30$^{+0.07+0.03}_{-0.08-0.03}$&
                            147$\pm$7 & 15$\pm$2 &  64$\pm$07$\pm$09 & 16.15$^{+0.07+0.08}_{-0.08-0.08}$&
                            174$\pm$5 & 11$\pm$1 &  70$\pm$04$\pm$11 & 16.36$^{+0.06+0.07}_{-0.08-0.10}$&
                            199$\pm$6 & 17$\pm$2 &  \phn95$\pm$07$\pm$16 & 16.45$^{+0.06+0.04}_{-0.08-0.04}$\\

\hi $\,\lambda$917.181 & 8 & \phn\phn90$\pm$14 & 29$\pm$3 &  48$\pm$23$\pm$07 & 16.02$^{+0.12+0.06}_{-0.17-0.07}$&
 145$\pm$6 & 15$\pm$2 &  70$\pm$08$\pm$07 & 16.32$^{+0.06+0.04}_{-0.07-0.04}$&
 173$\pm$5 & 10$\pm$1 &  68$\pm$05$\pm$11 & 16.44$^{+0.05+0.09}_{-0.05-0.11}$&
 197$\pm$7 & 15$\pm$3 &  \phn62$\pm$12$\pm$13 & 16.25$^{+0.05+0.07}_{-0.05-0.08}$\\

\cw $\,\lambda$1334.532 & 9 & 104$\pm$7 & 16$\pm$3 & 52$\pm$06$\pm$04 & 13.52$^{+0.05+0.02}_{-0.07-0.03}$& 
 145$\pm$3 & 13$\pm$1 & 91$\pm$05$\pm$03 & 13.90$^{+0.07+0.34}_{-0.09-0.01}$& 
 175$\pm$9 & 10$\pm$3 & 56$\pm$04$\pm$03 & 13.63$^{+0.04+0.05}_{-0.05-0.07}$ & 
 198$\pm$5 & 14$\pm$3 &101$\pm$06$\pm$14 & 13.92$^{+0.04+0.25}_{-0.05-0.02}$ \\

\cw $\,\lambda$1036.337 & 22 & 107$\pm$7 & 18$\pm$4 & 33$\pm$04$\pm$07 & 13.50$^{+0.05+0.06}_{-0.06-0.06}$& 
 146$\pm$6 & 15$\pm$2 & 52$\pm$03$\pm$07 & 13.77$^{+0.03+0.06}_{-0.03-0.07}$ &
 \nodata & \nodata & \tm{f} & \tm{f} &
 \nodata & \nodata & \tm{f} & \tm{f} \\

\ct $\,\lambda$977.020 & 8 & 105$\pm$8 & 19$\pm$6 & 83$\pm$06$\pm$09 & 13.28$^{+0.03+0.05}_{-0.02-0.05}$&
 146$\pm$6 & 12$\pm$6 & 93$\pm$04$\pm$13 & 13.59$^{+0.04+0.70}_{-0.03-0.23}$& 
 174$\pm$6 & 10$\pm$6 & 89$\pm$03$\pm$17 & 13.92$^{+0.03+1.00}_{-0.03-0.23}$& 
 197$\pm$6 & 15$\pm$6 &123$\pm$05$\pm$12 & 13.90$^{+0.03+1.00}_{-0.03-0.07}$ \\

\cf $\,\lambda$1550.770 & 9 & \nodata & \nodata & $<$45 & $<$13.35 & 
 150$\pm$7 & 14$\pm$5 & 37$\pm$10$\pm$11 & 13.36$^{+0.14+0.05}_{-0.20-0.05}$& 
 167$\pm$7 & 8$\pm$2 & 25$\pm$09$\pm$08 & 13.20$^{+0.13+0.04}_{-0.18-0.05}$ & 
 213$\pm$6 & 13$\pm$4 & \phn35$\pm$12$\pm$02 & 13.31$^{+0.17+0.02}_{-0.28-0.02}$\\

\ion{N}{1} $\,\lambda$1200.710 & 6 & \nodata & \nodata & $<$43 & $<$13.88&
 \nodata & \nodata & $<$21 & $<$13.58& 
 \nodata & \nodata & $<$26 & $<$13.66& 
 \nodata & \nodata & $<$34 & $<$13.78 \\

\ion{N}{2} $\,\lambda$1083.994& 6 & \nodata & \nodata & $<$55 & $<$13.68 &
 \nodata & \nodata & $<$42 & $<$13.57 &
 \nodata & \nodata & $<$33 & $<$13.46 &
 \nodata & \nodata & $<$40 & $<$13.55 \\

\nf $\,\lambda$1238.821& 10 & \nodata & \nodata & $<$44 & $<$13.32 & 
 \nodata & \nodata & $<$21 & $<$13.00& 
 \nodata & \nodata & $<$20 & $<$12.97 & 
 \nodata & \nodata & $<$47 & $<$13.35 \\

\oi $\,\lambda$1302.169  & 12 & \nodata & \nodata & $<$23 & $<$13.48 & 
 \nodata & \nodata & $<$29 & $<$13.58 & 
 \nodata & \nodata & $<$16 & $<$13.33 & 
 \nodata & \nodata & $<$18 & $<$13.37 \\

\os $\,\lambda$1031.926  & 22 &  \phn90$\pm$3 & 33$\pm$2 & 21$\pm$04$\pm$03 & 13.25$^{+0.10+0.07}_{-0.11-0.07}$& 
 145$\pm$6 & 15$\pm$1 & 26$\pm$03$\pm$03 & 13.36$^{+0.06+0.08}_{-0.06-0.09}$ &
 173$\pm$6 & 11$\pm$1 & 23$\pm$03$\pm$06 & 13.34$^{+0.05+0.06}_{-0.06-0.07}$& 
 201$\pm$7 & 19$\pm$2 & \phn22$\pm$04$\pm$04 & 13.29$^{+0.07+0.05}_{-0.09-0.06}$\\
% \nodata & \nodata  & \tm{f} & \tm{f} &
% \nodata & \nodata  & \tm{f} & \tm{f} \\ 
%$>$19 & $>$13.16\\

\os $\,\lambda$1037.617  & 24 & \nodata & \nodata & $<$20 & $<$13.50&
 \nodata & \nodata & \tm{f} & \tm{f} &
 \nodata & \nodata & \tm{f} & \tm{f} &
 \nodata & \nodata & \tm{f} & \tm{f} \\

\siw $\,\lambda$1260.422 & 9 & 102$\pm$7 & 18$\pm$3 & 47$\pm$08$\pm$04 & 12.58$^{+0.10+0.02}_{-0.12-0.01}$& 
  145$\pm$3 & 12$\pm$2 & 62$\pm$06$\pm$04 & 12.81$^{+0.07+0.02}_{-0.08-0.02}$& 
  174$\pm$8 & 11$\pm$2 & 42$\pm$05$\pm$03 & 12.58$^{+0.07+0.04}_{-0.09-0.05}$& 
  199$\pm$6 & 13$\pm$3 & \phn51$\pm$07$\pm$11 & 12.64$^{+0.03+0.08}_{-0.03-0.11}$\\

\siw $\,\lambda$1193.290 & 6 & 101$\pm$9 & 21$\pm$6 & 27$\pm$12$\pm$04 & 12.70$^{+0.20+0.03}_{-0.37-0.03}$&  
  145$\pm$4 & 12$\pm$3 & 34$\pm$10$\pm$03 & 12.78$^{+0.15+0.02}_{-0.23-0.01}$&  
  \nodata & \nodata & $<$42 & $<$12.76 & 
  205$\pm$5 & 18$\pm$5 & \phn25$\pm$12$\pm$07 & 12.72$^{+0.14+0.02}_{-0.21-0.02}$ \\

\siw $\,\lambda$1190.416 & 6 & 105$\pm$6 & 15$\pm$4 & 32$\pm$13$\pm$03 & 13.05$^{+0.19+0.03}_{-0.35-0.03}$ & 
   138$\pm$7 & 10$\pm$3 & 22$\pm$11$\pm$09 & 12.89$^{+0.18+0.04}_{-0.29-0.04}$ & 
   176$\pm$6 &  8$\pm$3 & 29$\pm$09$\pm$09 & 12.99$^{+0.13+0.02}_{-0.20-0.03}$ & 
   \nodata & \nodata &$<$60 &$<$13.21 \\

\siw $\,\lambda$1526.710 & 10 & \nodata & \nodata & $<$17 &  $<$12.88 & 
   \nodata & \nodata & $<$42 & $<$13.27 & 
   \nodata & \nodata &  $<$14 & $<$12.80 & 
   206$\pm$5 & 18$\pm$4 & \phn24$\pm$08$\pm$01 & 13.08$^{+0.18+0.02}_{-0.29-0.02}$ \\

\sit $\,\lambda$1206.500 & 6 &  \phn88$\pm$7 & $9\pm$4 & 25$\pm$08$\pm$08 & 12.22$^{+0.16+0.02}_{-0.26-0.02}$ & 
  146$\pm$5 & 14$\pm$2 & 79$\pm$07$\pm$06 & 12.76$^{+0.07+0.02}_{-0.08-0.02}$ & 
  173$\pm$8 & 10$\pm$1 & 63$\pm$06$\pm$07 & 12.67$^{+0.08+0.04}_{-0.09-0.03}$ & 
  202$\pm$5 & 13$\pm$3 &100$\pm$08$\pm$10 & 12.89$^{+0.01+0.28}_{-0.01-0.07}$ \\

\sif $\,\lambda$1393.755 & 10 & \nodata & \nodata & $<$34 & $<$12.57& 
   \nodata & \nodata & $<$30 & $<$12.52& 
   \nodata & \nodata & $<$21 & $<$12.36& 
   202$\pm$3 & 10$\pm$5 & \phn37$\pm$07$\pm$05 & 12.70$^{+0.11+0.02}_{-0.15-0.02}$  \\

\sif $\,\lambda$1402.770 & 12 & 
 \nodata & \nodata & $<$21 & $<$12.67 &
 \nodata & \nodata & $<$23 & $<$12.72 &
 \nodata & \nodata & $<$37 & $<$12.92 &
  200$\pm$5 & 18$\pm$2 & \phn38$\pm$09$\pm$04 & 13.00$^{+0.12+0.02}_{-0.15-0.02}$ \\

\ion{P}{2} $\,\lambda$1152.818& 15 & \nodata & \nodata & $<$34 & $<$13.09&
 \nodata & \nodata & $<$25 & $<$12.95&
 \nodata & \nodata & $<$23 & $<$12.92&
 \nodata & \nodata & $<$16 & $<$12.77 \\

\ion{S}{2} $\,\lambda$1259.519& 10 & \nodata & \nodata & $<$34 & $<$14.17& 
 \nodata & \nodata & $<$32 & $<$14.14 &
 \nodata & \nodata & \tm{f}&\tm{f} &
 \nodata & \nodata & \tm{f}&\tm{f}\\

\ion{S}{2} $\,\lambda$1253.811 & 11 & \nodata & \nodata & $<$26 & $<$14.23&
 \nodata & \nodata & $<$25 & $<$14.22 &
 \nodata & \nodata & $<$20  & $<$14.12& 
 \nodata & \nodata & $<$21 & $<$14.14\\

\ion{S}{3} $\,\lambda$1012.495& 20 & \nodata & \nodata & $<$20 & $<$13.71 & 
 \nodata & \nodata & $<$12 & $<$13.48 & 
 \nodata & \nodata & $<$15 & $<$13.58 &
 \nodata & \nodata & $<$17 & $<$13.66 \\

\ion{Ar}{1} $\,\lambda$1048.220& 26 & \nodata & \nodata & $<$16 & $<$12.81 &
 \nodata & \nodata & $<$12 & $<$12.68 & 
 \nodata & \nodata & $<$18 & $<$12.86 & 
 \nodata & \nodata & $<$14 & $<$12.76 \\

\few $\,\lambda$1144.938 & 16 & \nodata & \nodata & $<$17 & $<$13.14 & 
 145$\pm$2 & 8$\pm$3 & 11$\pm$4$\pm$5 & 13.01$^{+0.10+0.14}_{-0.12-0.21}$ & 
 \nodata & \nodata & $<$10 & $<$12.88 &
 \nodata & \nodata & $<$13 & $<$13.02 \\

\few $\,\lambda$1608.451& 6 & \nodata & \nodata & $<$91 & $<$13.81 &
 \nodata & \nodata & $<$45 & $<$13.50 &
 \nodata & \nodata & $<$41 & $<$13.46 &
 \nodata & \nodata & $<$75 & $<$13.72 \\

\ion{Fe}{3} $\,\lambda$1122.524 & 17 & \nodata & \nodata & $<$17 & $<$13.27 &
 \nodata & \nodata & \tm{f}&\tm{f} &
 \nodata & \nodata & \tm{f}&\tm{f} &
 \nodata & \nodata & \tm{f}&\tm{f} \\
\enddata

\tn{a}{Signal-to-noise per resolution element in the continuum next
  to the line.}
\tn{b}{$\bar{v}=\int^{v_+}_{v_-}v\tau_a(v)\mathrm{d}v/\int^{v_+}_{v_-}\tau_a(v)\mathrm{d}v$
  (\kms).}
\tn{c}{$b=\sqrt{2\int^{v_+}_{v_-}(v-\bar{v})^2\tau_a(v)\mathrm{d}v/\int^{v_+}_{v_-}\tau_a(v)\mathrm{d}v}$.
  (\kms).}
\tn{d}{Equivalent width measured between quoted velocity limits
  (m\AA), with statistical and systematic errors.}
\tn{e}{Apparent column density between quoted velocity
  limits, with statistical and systematic errors, or, for
  non-detections, 3$\sigma$ limit set on 
  the column density using 3$\sigma$ equivalent width limit and linear
  COG. $N_a$ is in\sqcm.}
\tn{f}{No measurement or upper limit possible, because of blending
  with other lines. We list here the contaminated line, contaminated components, blending line
  and velocity:  
  \hi $\,\lambda919.351$, HV1, \oi$\lambda919.658$ at $v\!\approx\!0$;
  \cw $\,\lambda1036.337$, HV3-4, \cw$^*$ $\lambda1037.018$ at $v\!\approx\!0$;
  \cf $\,\lambda1548.195$, HV1-4, \ion{N}{1} $\lambda1200.223$ at
  $v\!\approx\!0$; 
  \ion{N}{1} $\,\lambda1200.223$, HV1-4, \ion{N}{1} $\lambda1200.710$
  at $v\!\approx\!0$; 
  %\os $\,\lambda1031.926$, HV3-4, IGM \ion{Ne}{8} $\lambda770.409$ at
  %$z=0.34028$; 
  \os $\,\lambda1037.617$, HV2-4, H$_2$ $\lambda1038.156$ at $v\!\approx\!0$; 
  %\ion{S}{2} $\,\lambda$1250.584, HV1-4, IGM Ly$\beta$ at $z=0.22003$; 
  \ion{S}{2} $\,\lambda$1259.519, HV3-4, \siw$\,\lambda$1260.422 at
  $v\!\approx\!0$;   
  \ion{Fe}{3} $\,\lambda$1122.524, HV2-4, IGM Ly$\zeta$ at $z=0.20701$.} 
\end{deluxetable}
\end{turnpage}

\begin{deluxetable}{lcccc ccccc}
\tablewidth{0pt}
\tabcolsep=2pt
\tabletypesize{\scriptsize}
\tablecaption{Measurements of HVC Absorption toward \pg}
\tablehead{Line & S/N\tm{a} 
& \multicolumn{4}{c}{\underline{\phm{aaa}HV1, $-$165 to $-$125\kms}\phm{aaa}}
& \multicolumn{4}{c}{\underline{\phm{aaa}HV2, 80 to 160\kms\tm{g}}\phm{aaa}}\\
&& $\bar{v}$\tm{b} & $b$\tm{c} & $W_{\lambda}$\tm{d} & \phm{aaa}log\,$N_a$\tm{e}
& $\bar{v}$\tm{b} & $b$\tm{c} & $W_{\lambda}$\tm{d} & \phm{aaa}log\,$N_a$\tm{e}}

\startdata
\hi $\,\lambda$920.963 & 9 & $-$153$\pm$8 & 19$\pm$4 & 89$\pm$07$\pm$10 & 16.09$^{+0.02+0.05}_{-0.03-0.07}$&
 114$\pm$6  & 23$\pm$3 &125$\pm$09$\pm$08\phn & 16.26$^{+0.04+0.01}_{-0.06-0.02}$\\

\hi $\,\lambda$919.351 & 9 & $-$149$\pm$6 & 16$\pm$4 & 74$\pm$07$\pm$10 & 16.13$^{+0.04+0.07}_{-0.03-0.07}$&
 \nodata & \nodata & \tm{f} & \tm{f} \\
 %110$\pm$6  & 28$\pm$2 &173$\pm$09$\pm$19 & 16.58$^{+0.03+0.06}_{-0.03-0.07}$\\

\hi $\,\lambda$918.129 & 9 & $-$138$\pm$8& 16$\pm$2 & 77$\pm$08$\pm$10 & 16.24$^{+0.03+0.07}_{-0.03-0.08}$&
 109$\pm$7 & 14$\pm$8 & 73$\pm$12$\pm$08 & 16.20$^{+0.08+0.02}_{-0.10-0.02}$\\

\hi $\,\lambda$917.181 & 9 & \nodata & \nodata & \tm{f}&\tm{f} &
 \phn118$\pm$10  & 24$\pm$3 & 84$\pm$11$\pm$09 & 16.32$^{+0.08+0.03}_{-0.09-0.02}$\\

\cw $\,\lambda$1334.532 & 12 & $-$147$\pm$6 & 12$\pm$4 & 54$\pm$05$\pm$02 & 13.57$^{+0.05+0.02}_{-0.05-0.02}$ & 
 116$\pm$6 & 15$\pm$7 & 49$\pm$09$\pm$03 &13.48$^{+0.07+0.01}_{-0.08-0.01}$ \\

\cw $\,\lambda$1036.337 & 12 & \nodata & \nodata & $<$38 & $<$13.53 &
 \nodata & \nodata & $<$46 & $<$13.59\\

\ct $\,\lambda$977.020  & 13 & \nodata & \nodata & \tm{f}&\tm{f} &
 127$\pm$7 & 40$\pm$5 & 25$\pm$14$\pm$16 & 13.77$^{+0.02+0.04}_{-0.02-0.04}$\\

\cf $\,\lambda$1548.195 & 9 &\nodata & \nodata & $<$38 & $<$12.97 &
 153$\pm$6 & 23$\pm$4 & 35$\pm$13$\pm$01 & 12.99$^{+0.13+0.01}_{-0.20-0.01}$\\

\cf $\,\lambda$1550.770 & 8 & \nodata & \nodata & $<$47 & $<$13.36 &
 \nodata & \nodata & $<$38 & $<$13.27\\

\ion{N}{1} $\lambda$1199.550 & 7 &\nodata & \nodata & $<$45 & $<$13.42&
 \nodata & \nodata & \tm{f}&\tm{f}\\

\ion{N}{1} $\,\lambda$1200.710 & 7 & \nodata & \nodata & \tm{f}&\tm{f} &
 \nodata & \nodata & $<$86 & $<$14.19\\

\ion{N}{2} $\,\lambda$1083.994  & 6 & \nodata & \nodata & $<$58 & $<$13.71&
 \nodata & \nodata & $<$101 & $<$13.95\\ 

\nf $\,\lambda$1238.821 & 9 & \nodata &\nodata & $<$22 & $<$13.01&
 \nodata & \nodata & $<$45 & $<$13.32\\

\oi $\,\lambda$1302.169 & 12 & \nodata &\nodata & $<$19 & $<$13.39&
 \phn113$\pm$10 & 33$\pm$6 & 23$\pm$08$\pm$02 & 13.53$^{+0.12+0.03}_{-0.17-0.03}$\\

\os $\,\lambda$1031.926 & 14 & \nodata & \nodata & $<$39 & $<$13.49&
 \phn110$\pm$11 & 37$\pm$5 & 60$\pm$07$\pm$10 & 13.74$^{+0.05+0.06}_{-0.05-0.06}$\\

\siw $\,\lambda$1260.422 & 12 & $-$147$\pm$7 & 14$\pm$4 & 44$\pm$04$\pm$02 & 12.57$^{+0.05+0.01}_{-0.05-0.01}$&
 \nodata & \nodata & $<$43 & $<$12.45\\ 

\siw $\,\lambda$1193.290 & 6 & \nodata & \nodata & $<$38 & $<$12.71& 
 \nodata & \nodata & $<$63 & $<$12.93 \\
\siw $\,\lambda$1190.416 & 6 & \nodata & \nodata & $<$61 & $<$13.22&
 \nodata & \nodata & $<$44 & $<$13.07 \\
\siw $\,\lambda$1526.710 & 14 & \nodata & \nodata & $<$29 & $<$13.10&
 \nodata & \nodata & $<$24 & $<$13.02 \\

\sit $\,\lambda$1206.500 & 9 & $-$144$\pm$2 & 12$\pm$4 & 55$\pm$07$\pm$08 & 12.55$^{+0.03+0.06}_{-0.02-0.07}$& 
 128$\pm$8 & 42$\pm$4 & 30$\pm$14$\pm$11 & 12.90$^{+0.05+0.01}_{-0.05-0.01}$\\

\sif $\,\lambda$1393.755 & 11 & $-$144$\pm$7 & 15$\pm$3 & 20$\pm$07$\pm$03 & 12.39$^{+0.14+0.03}_{-0.19-0.02}$&
 150$\pm$6 & 14$\pm$6 & 18$\pm$08$\pm$02 & 12.34$^{+0.16+0.02}_{-0.24-0.01}$\\

\sif $\,\lambda$1402.770 & 9 & \nodata & \nodata & $<$25 & $<$12.74&
 \nodata & \nodata & $<$36 & $<$12.90\\ 

\ion{P}{2} $\,\lambda$1152.818& 17 &\nodata & \nodata & $<$27 & $<$12.99&
 \nodata & \nodata & $<$42 & $<$13.18\\

\ion{S}{2} $\,\lambda$1259.519 & 12 & \nodata & \nodata & $<$12 & $<$13.74 &
 \nodata & \nodata & $<$03 & $<$13.52\\

\ion{S}{3} $\,\lambda$1012.495 & 16 & \nodata & \nodata & $<$43 & $<$14.05 &
 \nodata & \nodata & \tm{f}&\tm{f}\\

\ion{Ar}{1} $\,\lambda$1048.220& 18 &\nodata & \nodata & $<$26 & $<$13.02& 
 \nodata & \nodata & $<$26 & $<$13.02\\

\few $\,\lambda$1144.938   & 18 & \nodata & \nodata & $<$35 & $<$13.44&
 \nodata & \nodata & $<$34 & $<$13.13\\ 

\ion{Fe}{3} $\,\lambda$1122.524 & 18 & \nodata & \nodata & \tm{f}&\tm{f}&
 \nodata & \nodata & $<$47 & $<$13.72\\

\enddata
\tn{a}{Signal-to-noise per resolution element in the continuum next
  to the line.}
\tn{b}{$\bar{v}=\int^{v_+}_{v_-}v\tau_a(v)\mathrm{d}v/\int^{v_+}_{v_-}\tau_a(v)\mathrm{d}v$
  (\kms).}
\tn{c}{$b=\sqrt{2\int^{v_+}_{v_-}(v-\bar{v})^2\tau_a(v)\mathrm{d}v/\int^{v_+}_{v_-}\tau_a(v)\mathrm{d}v}$.
  (\kms).}
\tn{d}{Equivalent width measured between quoted velocity limits
  (m\AA), with statistical and systematic errors.} 
\tn{e}{Apparent column density between quoted velocity
  limits, with statistical and systematic errors, or, for non-detections, 3$\sigma$ limit set on
  the column density using 3$\sigma$ equivalent width limit and linear
  COG. $N_a$ is in\sqcm.}
\tn{f}{No measurement or upper limit possible, because of blending.
  We list here the contaminated line, contaminated components, blending line and velocity: 
  \hi $\,\lambda919.351$, HV2, \oi $\lambda$919.658 at $v\!\approx\!0$;
  \hi $\,\lambda917.181$, HV1, edge of detector;
  \ct $\,\lambda977.020$, HV1, H$_2$ $\,\lambda976.552$ at $v\!\approx\!0$;
  \ion{N}{1} $\,\lambda1199.550$, HV2, \ion{N}{1} $\,\lambda1200.223$
  at $v\!\approx\!0$; 
  \ion{N}{1} $\,\lambda1200.223$, HV1, \ion{N}{1} $\,\lambda1199.550$
  at $v\!\approx\!0$; 
  \ion{N}{1} $\,\lambda1200.223$, HV2, \ion{N}{1} $\,\lambda1200.710$
  at $v\!\approx\!0$; 
  \ion{N}{1} $\,\lambda1200.710$, HV1, \ion{N}{1} $\,\lambda1200.223$
  at $v\!\approx\!0$; 
  \os $\,\lambda1037.617$, HV1, H$_2$ $\,\lambda1037.146$ at $v\!\approx\!0$;
  \os $\,\lambda1037.617$, HV2, H$_2$ $\,\lambda1038.156$ at $v\!\approx\!0$;
  \ion{S}{3} $\,\lambda$1012.495, HV2, H$_2$ $\,\lambda1012.822$ at
  $v\!\approx\!0$;   
  \ion{Fe}{3} $\,\lambda$1122.524, HV1, \few$\,\lambda1121.975$ at
  $v\!\approx\!0$.}  
\tn{g}{Measurements of \ct, \sit, \ion{S}{3}, \ion{Fe}{3}, and \os\ in HV2 are made
  over the extended velocity range 80 to 200\kms; measurements of \cf,
  \nf, and \sif\ are made over 125 to 200\kms.}
\end{deluxetable}

\begin{deluxetable}{lcccc ccccc}
\tablewidth{0pt}
\tabletypesize{\scriptsize}
\tablecaption{Summary of HVC Column Density Measurements}
\tablehead{Sight Line & Comp. & \hi\ & \cw & \ct & \cf & \siw & \sit &
  \sif & \os}
\startdata
\he & HV1 & 16.29$\pm$0.05 & 13.53$_{-0.07}^{+0.05}$ & 13.28$_{-0.05}^{+0.06}$ & 
$<$13.35 & 12.65$_{-0.10}^{+0.10}$ & 12.22$_{-0.26}^{+0.16}$& $<$12.57
& 13.25$_{-0.12}^{+0.13}$\\
    & HV2 & 16.21$\pm$0.10 & 13.90$_{-0.09}^{+0.34}$ & 13.59$_{-0.16}^{+0.70}$ & 
13.36$_{-0.21}^{+0.15}$ & 12.81$_{-0.08}^{+0.25}$ &
12.76$_{-0.08}^{+0.07}$ & $<$12.52 & 13.36$_{-0.11}^{+0.10}$\\
    & HV3 & 16.74$\pm$0.20\tm{a} & 13.63$_{-0.07}^{+0.06}$ & 13.92$_{-0.23}^{+1.00}$ & 
13.20$_{-0.19}^{+0.14}$ & 12.70$_{-0.07}^{+0.07}$ &
12.67$_{-0.09}^{+0.09}$ & $<$12.36 & 13.34$_{-0.09}^{+0.08}$\\
    & HV4 & 16.34$\pm$0.10 & 13.92$_{-0.05}^{+0.25}$ & 13.90$_{-0.05}^{+1.00}$ &
13.31$_{-0.28}^{+0.17}$ & 12.71$_{-0.08}^{+0.08}$ & 12.89$_{-0.07}^{+0.28}$ & 
12.85$_{-0.15}^{+0.15}$ & 13.29$_{-0.11}^{+0.09}$\\
\pg & HV1 & 16.15$\pm$0.15 & 13.57$_{-0.05}^{+0.05}$ & \nodata & $<$12.97 & 
12.57$_{-0.05}^{+0.05}$ & 12.55$_{-0.07}^{+0.07}$ &
12.39$_{-0.19}^{+0.14}$ & $<$13.49\\
    & HV2 & 16.26$\pm$0.15 & 13.48$_{-0.08}^{+0.07}$ & 13.77$_{-0.04}^{+0.25}$ & 
12.99$_{-0.20}^{+0.13}$ & $<$12.45 & 12.90$_{-0.05}^{+0.05}$ &
12.34$_{-0.24}^{+0.16}$ & 13.74$_{-0.08}^{+0.08}$\\
\enddata
\tablecomments{All measurements in this table are the logarithm of the
  ionic column density in cm$^{-2}$. Errors are $\pm$1$\sigma$; 
  %we include a 0.25\,dex allowance for saturation where appropriate; 
  upper/lower limits are 3$\sigma$. For cases where absorption column
  densities are measured in more than one line 
  (particularly \siw), we considered the dispersion of individual
  measurements and their errors to determine the final value listed.}
\tn{a}{Component saturated, even in weakest Lyman lines; column
  density estimated using AOD 
  correction method for saturated lines \citep{SS91}.}
\end{deluxetable}

\begin{deluxetable}{lcccc c}
%\tablewidth{200pt}
\tablewidth{0pt}
\tabletypesize{\small}
\tablecaption{Flux of Ionizing Radiation near the Galaxy}
\tablehead{Sight Line & $b$ & Model\tm{a} & Distance
& log\,$\Phi_{<912}$\tm{b} & log\,$\Phi_{>912}$\tm{b} \\
& (\degr) & & (kpc) & ($\Phi$ in cm$^{-2}$\,s$^{-1}$)
& ($\Phi$ in cm$^{-2}$\,s$^{-1}$)}
\startdata
\he & $-$66 & MW  & 10    & 5.0 & 7.0\\
    & & & 50    & 4.5 & 6.5\\
    & & & 100   & 3.8 & 5.8\\
    & & EGB &\nodata& 4.0 & 4.8\\
\pg & +52 & MW  & 10    & 5.5 & 7.5\\
    & & & 50    & 4.7 & 6.6\\ 
    & & & 100   & 3.8 & 6.2\\
    & & EGB &\nodata& 4.0 & 4.8\\
\enddata
\tn{a}{MW = escaping Milky Way radiation (see \S6.2); EGB =
  extragalactic background radiation \citep{HM96}.}
\tn{b}{Total ionizing flux impinging on the clouds at
  hydrogen-ionizing (hard; $\lambda<912$\,\AA) and
  non-hydrogen-ionizing (soft; $2460$\,\AA\ $>\lambda>912$\,\AA) wavelengths. We assume
  $F_{\nu}=4\pi J_{\nu}$ when computing $\Phi$ for the EGB (see \S6.1).}
\end{deluxetable}

\begin{deluxetable}{lcccc ccccc ccc }
\tablewidth{0pt}
\tabcolsep=4pt
\tabletypesize{\scriptsize}
\tablecaption{Results of HVC Photoionization Modeling}
\tablehead{Sight Line & Comp. & $\bar{v}$ & Model & Dist. 
 & log\,$U$\tm{a} & [Z/H]\tm{b} & log\,$n_{\mathrm{H}}$\tm{c} &
 $l$\tm{d} & $P/k$\tm{e} & $f$(\ion{H}{2})\tm{f} & $I$(\ha)\tm{g}\\
& & (km\,s$^{-1}$) & & (kpc) & & & ($n_{\mathrm{H}}$ in cm$^{-3}$) & (pc) & (cm$^{-3}$\,K) & & (mR)}
\startdata
\he & HVC1 &   99 & MW+EGB &  10  & $-$3.75 & $-$0.6 & $-$1.7 &   11 &  470 & 0.98 & 2.2 \\
    &      &      &        &  50  & $-$3.70 & $-$0.6 & $-$2.1 &   39 &  170 & 0.98 & 0.8 \\
    &      &      &        & 100  & $-$3.80 & $-$0.6 & $-$2.4 &   56 &   79 & 0.97 & 0.3 \\
    &      &      &    EGB &\nodata &$-$3.80 & $-$0.5 & $-$2.6 &   86 &   51 & 0.97 & 0.2 \\
    & HVC2 &  148 & MW+EGB &  10  & $-$3.55 & $-$0.4 & $-$1.9 &   23 &  290 & 0.98 & 1.9 \\
    &      &      &        &  50  & $-$3.55 & $-$0.4 & $-$2.3 &   57 &  110 & 0.98 & 0.7 \\
    &      &      &        & 100  & $-$3.55 & $-$0.3 & $-$2.7 &  150 &   47 & 0.98 & 0.3 \\
    &      &      &    EGB &\nodata &$-$3.55 & $-$0.3 & $-$2.9 &  230 &   30 & 0.98 & 0.2 \\
    & HVC3 &  175 & MW+EGB &  10  & $-$3.75 & $-$0.9 & $-$1.7 &   37 &  550 & 0.98 & 6.3 \\
    &      &      &        &  50  & $-$3.70 & $-$0.9 & $-$2.1 &  120 &  190 & 0.98 & 2.3 \\
    &      &      &        & 100  & $-$3.70 & $-$0.9 & $-$2.5 &  290 &   78 & 0.98 & 0.9 \\
    &      &      &    EGB &\nodata &$-$3.70 & $-$0.9 & $-$2.7 &  440 &   50 & 0.98 & 0.5 \\
    & HVC4 &  193 & MW+EGB &  10  & $-$3.50 & $-$0.6 & $-$1.9 &   38 &  240 & 0.98 & 2.6 \\
    &      &      &        &  50  & $-$3.50 & $-$0.6 & $-$2.3 &   98 &   95 & 0.98 & 1.0 \\
    &      &      &        & 100  & $-$3.50 & $-$0.5 & $-$2.7 &  240 &   39 & 0.98 & 0.4 \\
    &      &      &    EGB &\nodata &$-$3.45 & $-$0.4 & $-$3.0 &  500 &   25 & 0.99 & 0.2 \\
\pg & HVC1 &$-$150& MW+EGB &  10  & $-$3.70 & $-$0.6 & $-$1.2 &    3 & 1180 & 0.98 & 5.2 \\
    &      &      &        &  50  & $-$3.65 & $-$0.6 & $-$2.0 &   24 &  200 & 0.98 & 0.9 \\
    &      &      &        & 100  & $-$3.66 & $-$0.6 & $-$2.5 &   81 &   59 & 0.98 & 0.3 \\
    &      &      &    EGB &\nodata &$-$3.60 & $-$0.5 & $-$2.8 &  170 &   36 & 0.98 & 0.1 \\
    & HVC2 &$-$125& MW+EGB &  10  & $-$3.15 & $-$0.9 & $-$1.8 &   56 &  460 & 1.00 & 5.0 \\
    &      &      &        &  50  & $-$3.15 & $-$0.8 & $-$2.5 &  310 &   86 & 1.00 & 0.9 \\
    &      &      &        & 100  & $-$3.11 & $-$0.8 & $-$3.1 & 1290 &   24 & 1.00 & 0.2 \\
    &      &      &    EGB &\nodata &$-$3.20 & $-$0.7 & $-$3.2 & 1270 &   18 & 0.99 & 0.1 \\

\enddata
\tablecomments{Our photoionization models are conducted by finding the
  values of the ionization parameter $U$ and metallicity [$Z$/H] that best reproduce
  the observed ionic column densities of \hi, \cw, \ct, \siw, and \sit, for
  a given input radiation field at a given distance.}
\tn{a}{Best fit ionization parameter, $\pm$0.2\,dex (95\% c.l.).}
\tn{b}{Best fit metallicity, $\pm$0.2\,dex (95\% c.l.), or
  $\pm$0.3\,dex for \he\ HVC3 and HVC4, where saturation may affect
  the \ct\ line.}
\tn{c}{Total hydrogen density:
  $n_{\mathrm{H}}=n_{\gamma}/U$\,cm$^{-3}$, where $n_{\gamma}$ is the density of
  ionizing photons in the radiation field.}
\tn{d}{Cloud depth: $l=N_{\mathrm{H}}/n_{\mathrm{H}}$, where
  $N_{\mathrm{H}}$ is the total hydrogen column density returned by
  the model.}
%\tn{e}{Cloud mass, in solar masses, assuming clouds are spherical; $M=(4/3)\pi r^3\mu
%  m_{\mathrm{H}}n_{\mathrm{H}}$, where the mass per hydrogen atom
%  $\mu=1.4$ and $r=l/2$.}
\tn{e}{HVC pressure, $P/k=\sum_in_iT$, summing over all elements and
  ionization states $i$ and their associated electrons.}
\tn{f}{Hydrogen ionization fraction.}
%\tn{h}{Angular size on the sky: $\theta=60\times(l/D)\times(180/\pi)$
%  arcminutes, assuming clouds are spherical.}
\tn{g}{Predicted intensity of \ha\
  emission: $I$(\ha)=2.75$(T/10^4\,\mathrm{K})^{0.924}n_e^2l$ 
  Rayleighs (R), where 1 R = $10^6/4\pi$\,photons\sqcm\,s$^{-1}$\,sr$^{-1}$.}
\end{deluxetable}

\begin{deluxetable}{lclll ll}
\tablewidth{0pt}
\tabletypesize{\footnotesize}
\tablecaption{High Ion Column Density Ratios in HVCs\tm{a}}
\tablehead{Sight Line & $\bar{v}$ & $\frac{N(\rm{H\;I})}{N(\rm{O\;VI})}$ &
  $\frac{N(\rm{Si\;IV})}{N(\rm{O\;VI})}$ & $\frac{N(\rm{C\;IV})}{N(\rm{O\;VI})}$ & 
  $\frac{N(\rm{N\;V})}{N(\rm{O\;VI})}$ &  $\frac{N(\rm{C\;IV})}{N(\rm{Si\;IV})}$\\
  & (\kms) & & & & &}
\startdata
\he  & +\phn99&$1100^{+300}_{-400}$& $<$0.12 & $<$0.72     & $<$0.80 & \nodata \\
     & +148   & $710^{+240}_{-230}$& $<$0.08 & $1.0\pm0.5$ & $<$0.26 & $>$10 \\
     & +175   &$2500^{+1500}_{-1100}$&$<$0.05& $0.7\pm0.3$ & $<$0.23 & $>$10 \\
     & +193   &$1100^{+400}_{-300}$&$0.36\pm0.16$ & $1.0\pm0.5$ & $<$0.65 & $2.9\pm1.8$\\
\pg  & $-$150 & $>$600            & $>$0.11 & \nodata     & \nodata & $<$2.6 \\
     & +124   & $330^{+150}_{-120}$  &$0.04\pm0.02$&$0.18\pm0.07$&$<$0.19& $4.5\pm2.5$\\
\tableline
PG~1259+593\tm{b} & $-110$  & $(1.7\pm0.3)\!\times\!10^6$ & $0.10\pm0.02$ & $0.35\pm0.05$ & $<0.07$ & $3.4\pm0.4$\\
Mrk~279\tm{c}     & $-140$  & $(6.8\pm1.6)\!\times\!10^5$ & $0.22\pm0.07$ & $0.89\pm0.29$ & $0.19\pm0.07$ & $4.0\pm1.6$ \\
PKS~2155--304\tm{d}& $-140$ & $370^{+250}_{-110}$         & $0.09\pm0.02$ & $0.51\pm0.06$ & $<0.08$ & $5.9\pm1.1$ \\
                   & $-170$ & $47^{+66}_{-19}$            & $0.05\pm0.02$ & $1.1\pm0.2$   & $<0.08$ & $1.1\pm0.2$ \\
Mrk~509\tm{d}      & $-240$ & $<2700$                     & $<0.02$       & $0.44\pm0.07$ & $<0.07$ & $>25$ \\
                   & $-300$ & $<2600$                     & $0.4\pm0.2$   & $1.7\pm0.4$   & $<0.09$ & $4.2\pm2.0$ \\
PG~1116+215\tm{e}  & +100   & $1350\pm40$                 & $0.18\pm0.06$ & $1.12\pm0.36$ & $<0.27$ & $6.3\pm1.3$ \\
                   & +184   & $7000^{+3000}_{-2000}$      & $0.12\pm0.01$ & $0.55\pm0.07$ & $<0.05$ & $4.5\pm0.8$ \\
\tableline
HVC average (detections only)\tm{f} & \nodata & \nodata & $0.17\pm0.13$ & $0.80\pm0.42$ & $0.19\pm0.07$ & $4.1\pm1.6$ \\
%HVC average (inc. limits)\tm{g} & \nodata & \nodata & $0.14\pm0.12$ & $0.73\pm0.44$ & $0.22\pm0.23$ & $6.0\pm6.2$ \\
Halo average\tm{g}& \nodata & \nodata & $0.20\pm0.13$ & $0.60\pm0.47$ & $0.12\pm0.07$ & $3.5\pm1.1$\\
\enddata
\tn{a}{Refer to Tables 3 and 4 for the velocity ranges and log~$N_a$
measurements for each component. Errors given are $\pm1\sigma$; upper
and lower limits are 3$\sigma$.} 
%\tn{b}{We only include the \os\ in the same velocity range as the \sif\
%  and \cf\ (125--200\kms) in the calculation of the high-ion ratios in
%  this component.}
\tn{b}{From \citet{Fo04}.}
\tn{c}{New measurements of $N$(\cf) and
  $N$(\sif) using archival spectra of Mrk~279. We measure
  $N$(\cf)=13.68$^{+0.11+0.02}_{-0.15-0.02}$ and
  $N$(\sif)=13.01$^{+0.10+0.01}_{-0.13-0.01}$ in the range
  $v_{LSR}$=$-$210 to $-$115\kms, where \hvo\ is detected.}
\tn{d}{From \citet{Co04}.}
\tn{e}{From \citet{Ga05}.}
\tn{f}{Mean value of detections, with quoted error representing
  sample standard deviation. Upper/lower limits are ignored.}
%\tn{g}{$1\sigma$ upper (or lower) limits treated as data point in calculation of
%  mean ionic ratio.}
\tn{g}{From \citet{Zs03}. Errors represent sample standard deviation.}
\end{deluxetable}

\begin{deluxetable}{lllll c}
\tablewidth{0pt}
\tabletypesize{\footnotesize}
\tabcolsep=5pt
\tablecaption{Origin Models for Highly Ionized HVC Components}
\tablehead{Origin & Ionization & 
  Explains\tm{a} & Explains\tm{a} & Explains\tm{a} & Ref.\\
& Source &
 $\frac{N(\rm{Si\;IV})}{N(\rm{O\;VI})}$? &
 $\frac{N(\rm{C\;IV})}{N(\rm{O\;VI})}$? &
 $\frac{N(\rm{N\;V})}{N(\rm{O\;VI})}$? &}
\startdata
Galactic Fountain & Recombination of hot gas& 0/8 (4/5) & \phn1/12 (1/1) & 0/1 (11/12) & (1) \\
HVC interface     & Conductive Heating      & 2/8 (4/5) & 11/12 (1/1) & 1/1 (11/12) & (2) \\
                  & Turbulent Mixing        & 7/8 (2/5) & \phn6/12 (0/1) & 1/1 (\phn6/12)  & (3) \\
                  & Shock Heating           & 6/8 (5/5) & 11/12 (1/1) & 0/1 (12/12) & (4) \\
\enddata
%\tn{a}{In this column we note whether the model predicts a kinematic
%  alignment between absorption lines of low and high ionization.}
\tn{a}{In these columns we assess what fraction of highly ionized HVCs
  ratios (from this paper and the literature) can be reproduced 
  (to within their $\pm1\sigma$ errors) by
  the given model assuming solar abundances. The numbers in parentheses
  represent the fraction of upper/lower limits that are consistent
  with the model predictions.  See \citet{Fo04} for a
  tabulation of ionic ratio predictions in different models.} 
\tablerefs{(1): \citet{EC86}; (2): \citet{Bo90}; (3): \citet{Sl93};
 (4): \citet{DS96}.}
\end{deluxetable} 

%\begin{figure}[!ht]
%\epsscale{0.6}
%\figurenum{1}
%\plotone{f1.eps}
%\caption{Map showing the distribution of \hvo\ detections (circles)
%  superimposed on the high-velocity \hi\ sky, and color-coded
%  according to velocity (S03). Galactic coordinates are
%  used with an equal-area Aitoff projection centered on $l=180\degr$. Note
%  that the sight lines toward \he\ and \pg\ lie in directions where no
%  high-velocity 21-cm \hi\ emission is seen.}
%\end{figure}

\begin{figure}[!ht]
\epsscale{1.0}
\plottwo{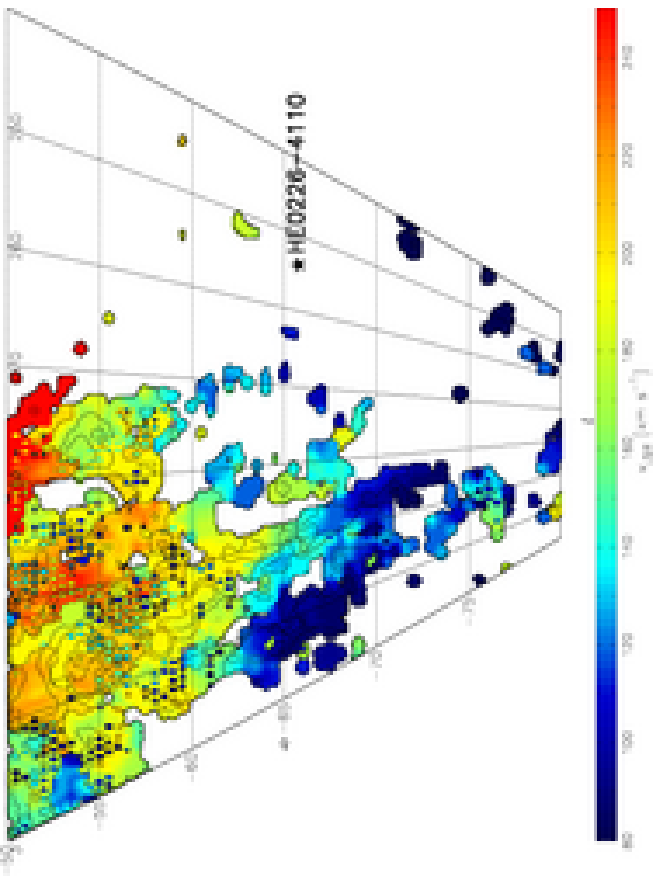}{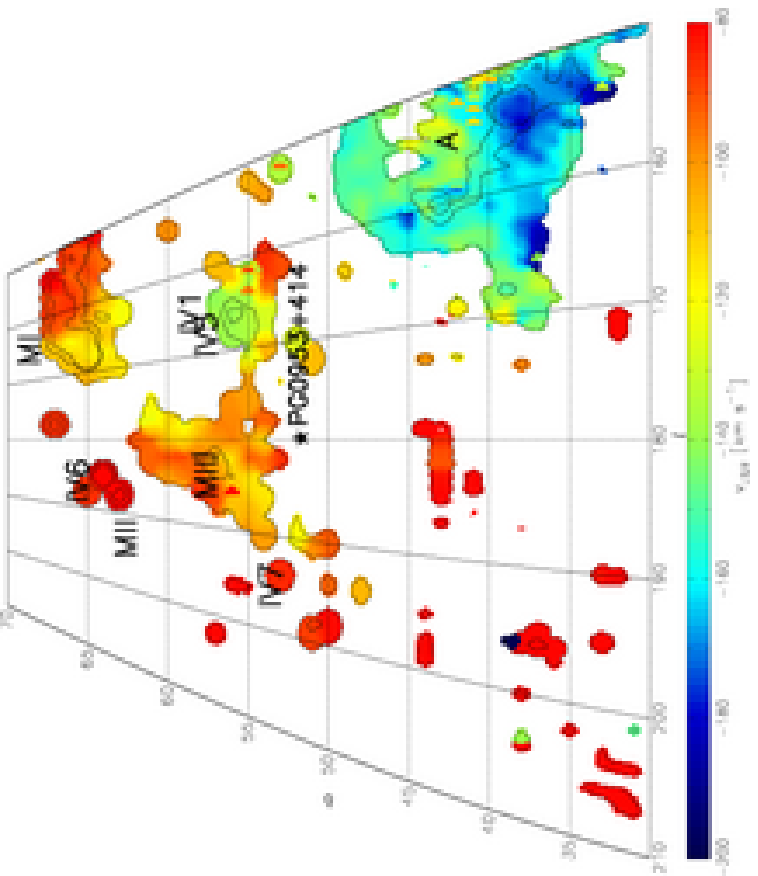}
\caption{Maps in Galactic coordinates ($l, b$) of the HVC velocity
  field in the regions around \he\ and \pg, based 
  on the \hi\ 21-cm data of \citet{HW88} and \citet{Mr00}.
  Colors show $v_{LSR}$, and contours are at brightness temperature levels of
  0.05, 0.5, and 1\,K, corresponding to column densities of about 2, 20, and
  40$\times10^{18}$\sqcm. Individual cloud cores 
  %in the Intermediate-Velocity Arch 
  are labeled in the \pg\ panel.}
\end{figure}

\clearpage
\begin{figure}[!ht]
\epsscale{0.85}
\plotone{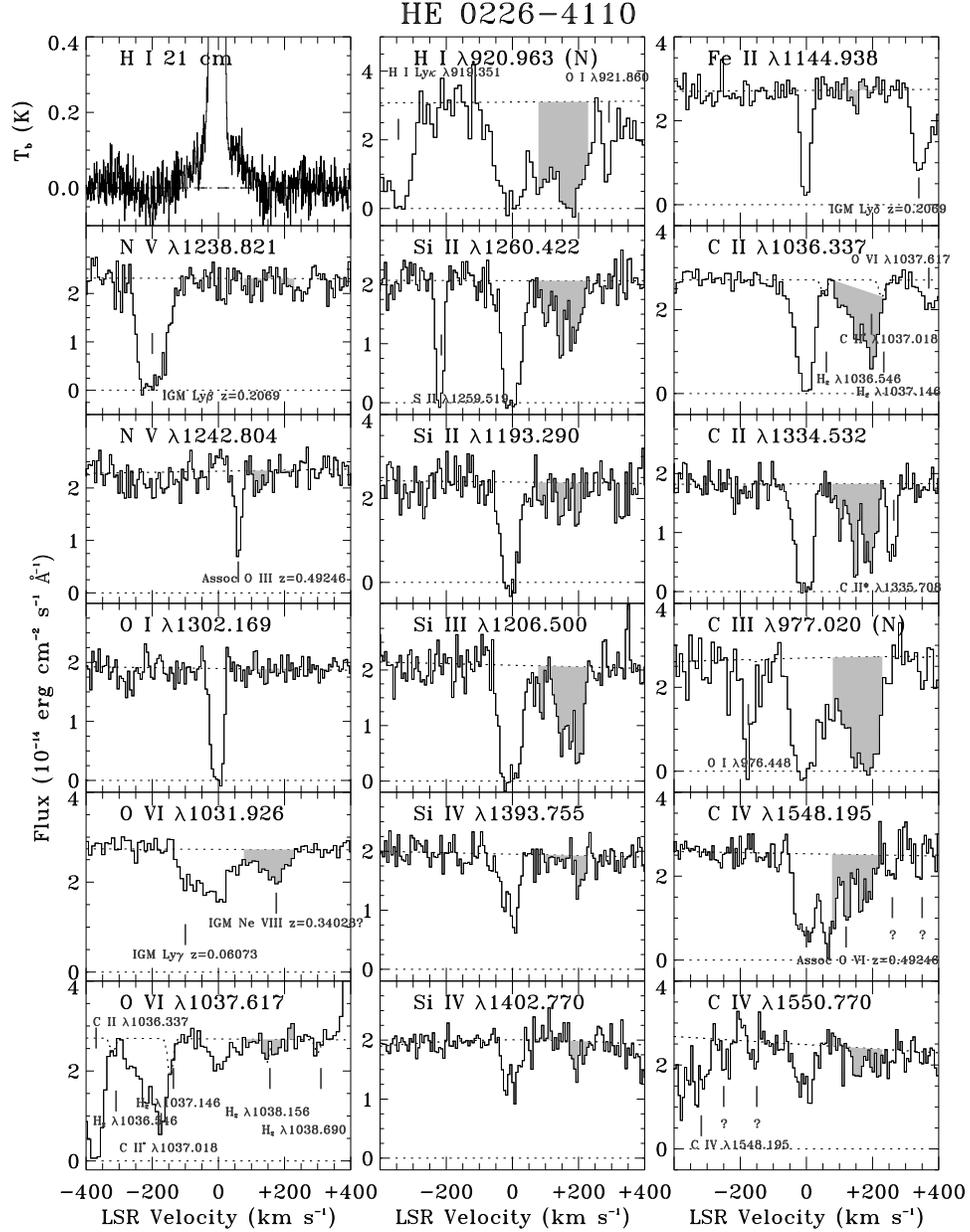}
\caption{Absorption line profiles along the line of sight to
  \he. Flux is plotted against LSR velocity with the continuum and
  zero levels shown
  as dotted lines. The upper left panel shows the 21-cm emission line
  profile. Absorption in the HVCs is shaded, and blends are
  annotated at the expected line center of the offending feature.
  For \os\ $\lambda1037.617$ and \cw\ $\lambda1036.337$, the continuum 
  includes a model of the H$_2$ lines. (N) indicates night-only data
  is displayed to reduce geocoronal airglow emission.}
\end{figure}

\begin{figure}[!ht]
\epsscale{0.85}
\plotone{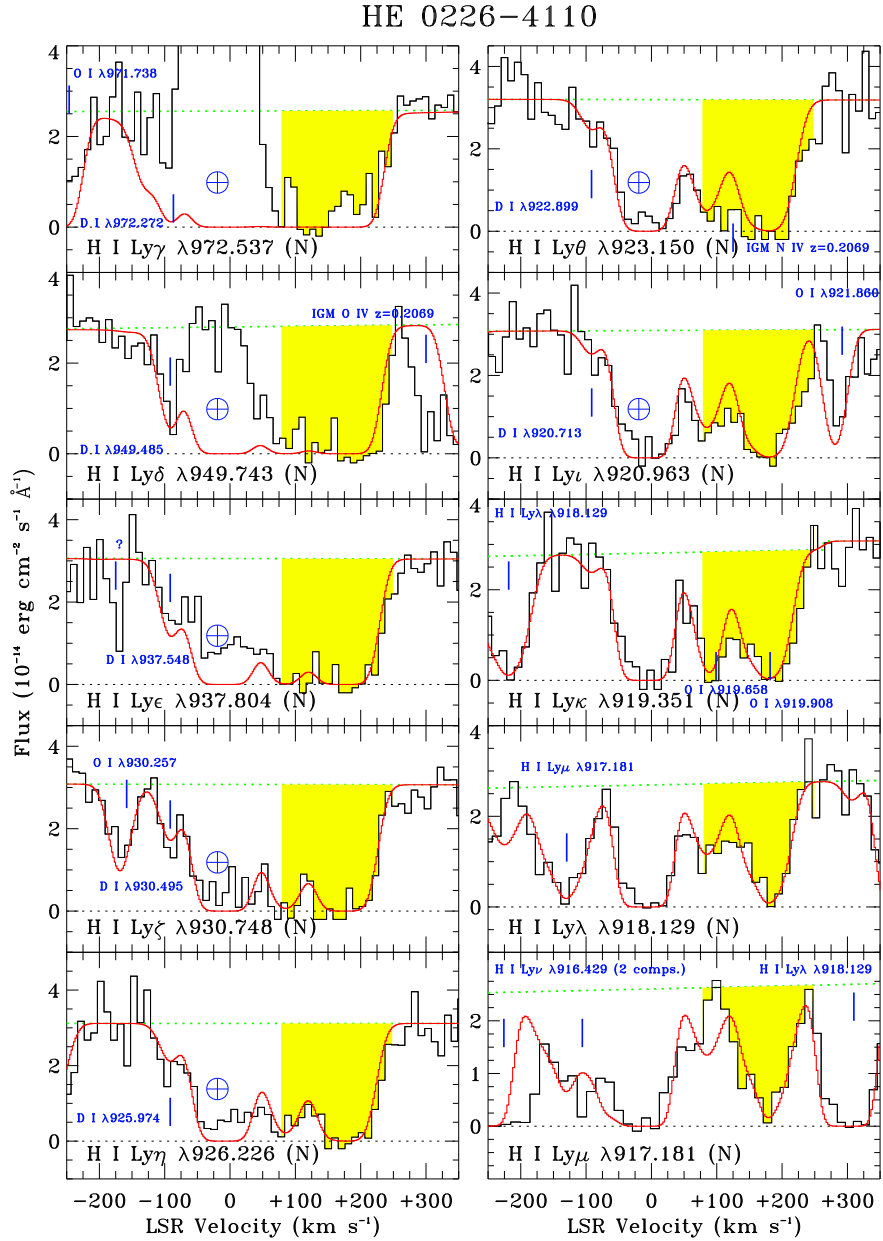}
\caption{\hi\ Lyman series absorption line profiles along the
  line of sight to \he. Dotted lines display the position of the continuum
  and the zero level. Geocoronal \hi\ emission centered near 0\kms\ is seen in the 
  stronger lines and can be ignored. The red line shows a Voigt
  component model of the \hi\ absorption,
  using the component velocities and column densities derived using the AOD
  method.} 
  %The model reproduces the data well with a small discrepancy
  %near 80\kms\ in the \hi\ $\lambda917.181$ line.} 
\end{figure}

\begin{figure}[!ht]
\epsscale{0.85}
\plotone{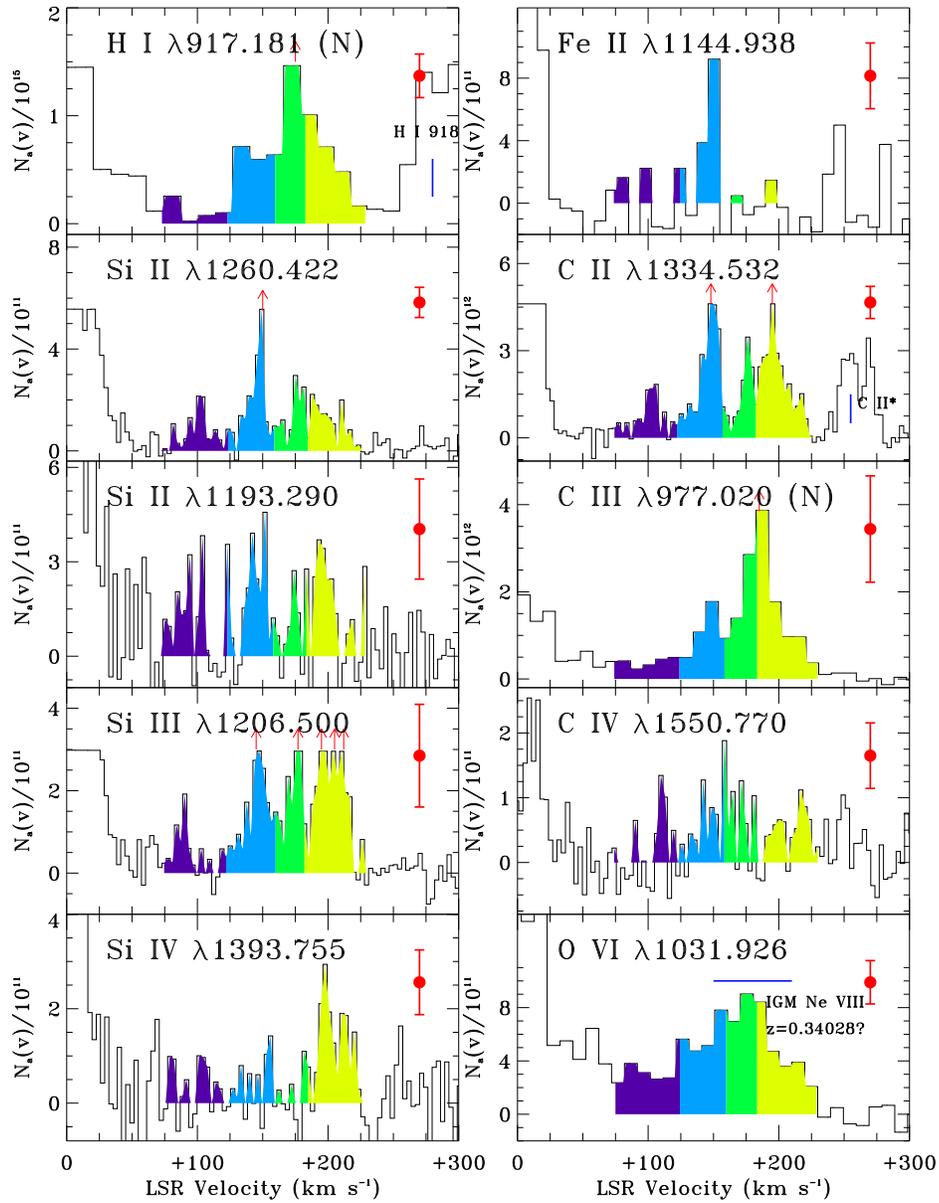}
\caption{Comparison of apparent column density as a function of
  velocity for various species seen in the HVCs
  toward \he. The units on the y-axes are ions\sqcm\
  (km\,s$^{-1}$)$^{-1}$. Error bars showing the typical uncertainty on
  each data point are shown in the corner of each panel, and small red
  arrows indicate where the high-velocity absorption may be saturated. 
  The color of the shading indicates the velocity range covered by
  each component.}
\end{figure}

\begin{figure}[!ht]
\epsscale{0.85}
\plotone{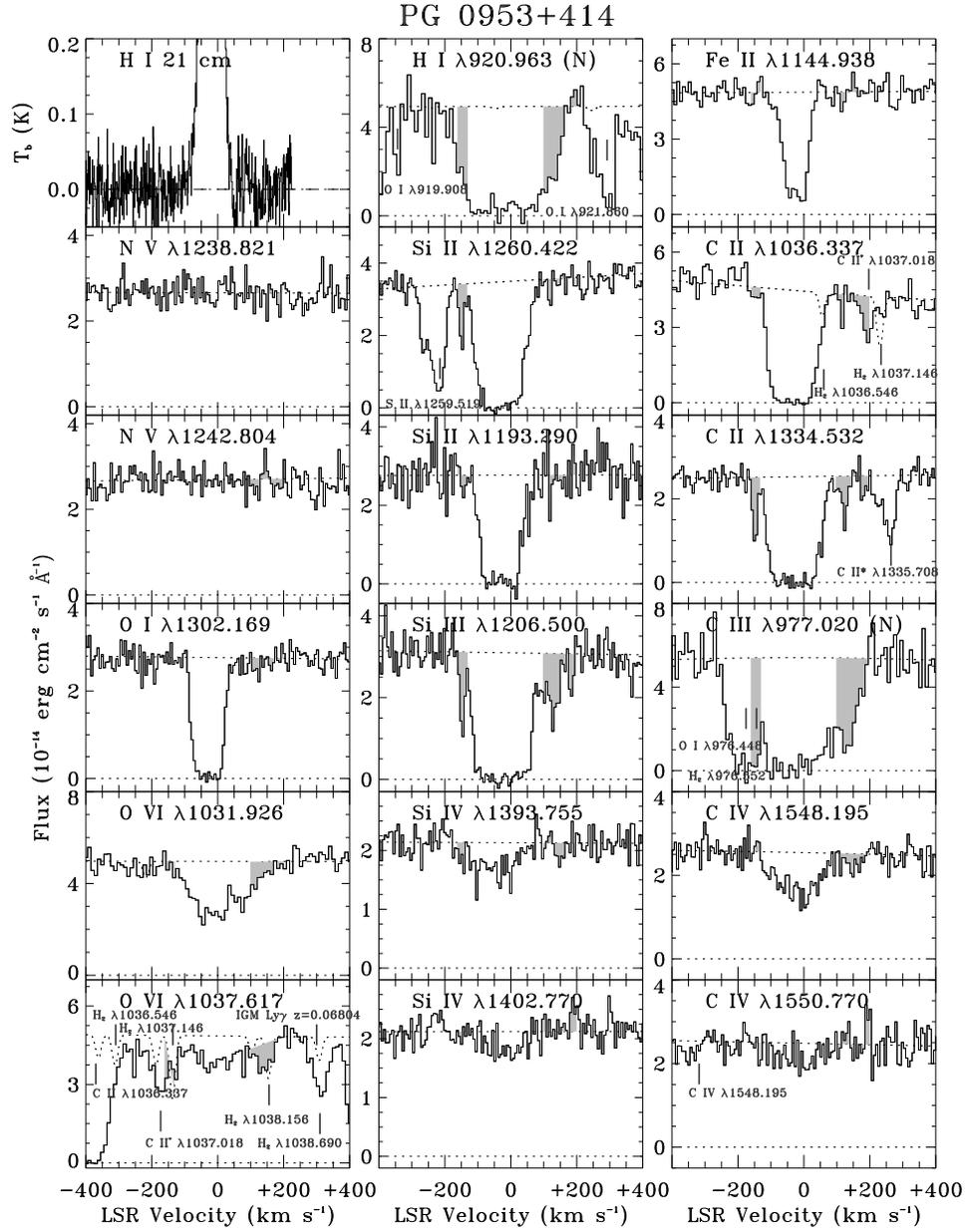}
\caption{Same as Figure 2, except for \pg.}
\end{figure}

\begin{figure}[!ht]
\epsscale{0.85}
\plotone{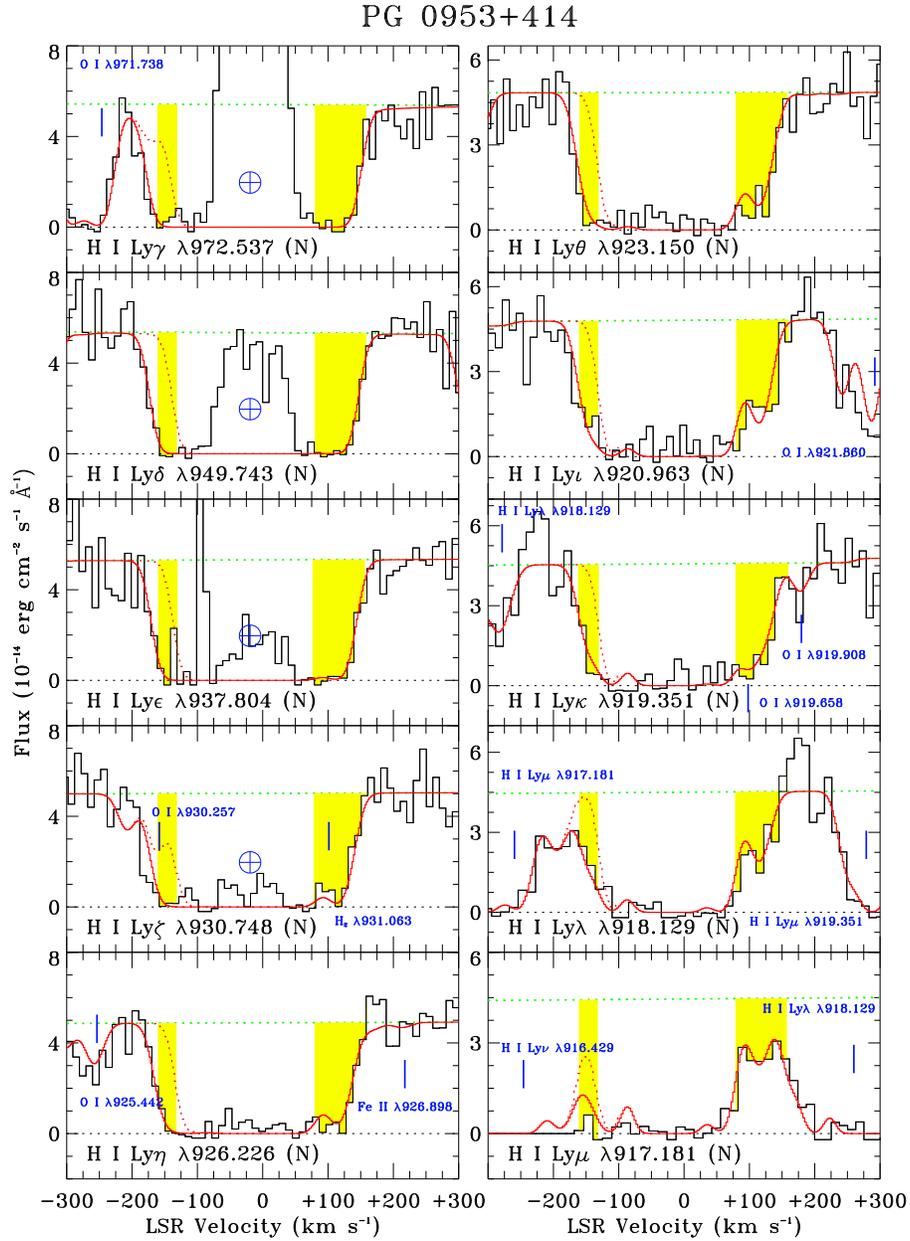}
\caption{Same as Figure 3, except for \pg. The dotted red line shows
  the Voigt profile model without a component at $-$150\kms; this model clearly does not
  match the data, showing that \hi\ exists and can be measured in the
  $-$150\kms\ HVC.}
  %There are small discrepancies near $-$200\kms\ in
  %the \hi\ $\lambda930.748$ line and near 120\kms\ in the \hi\
  %$\lambda920.963$ line.} 
%The \hi\ in the $-$150\kms\
%HVC is blended with low-velocity gas so we do not present \hi\ column density
%measurements in this absorber.}
\end{figure}

\begin{figure}[!ht]
\epsscale{0.85}
\plotone{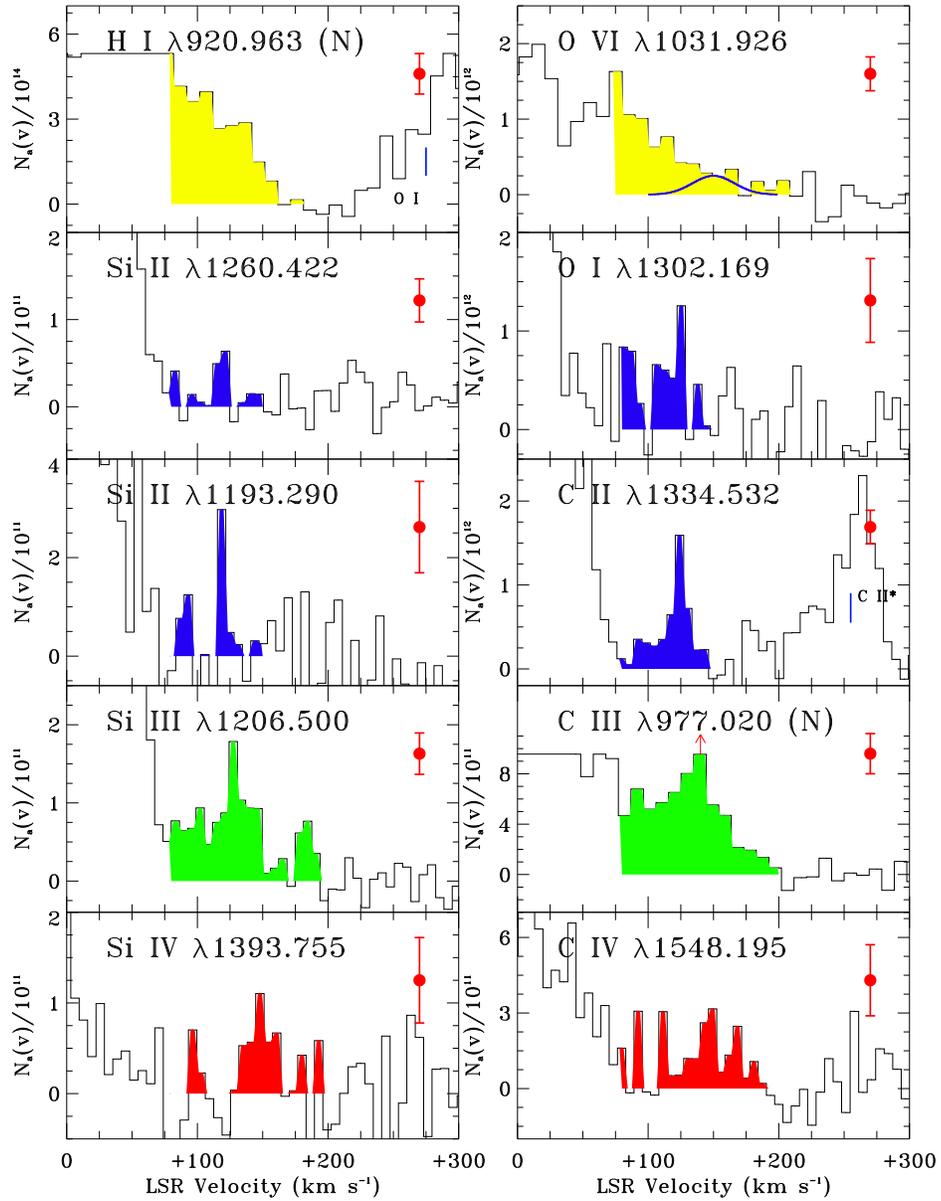}
\caption{Same as Figure 4, except for \pg. The color
  scheme reflects the shape of the absorption profile: narrow (blue), broad
  (green), broad and offset (red), and wing-like (yellow). The thick solid line
  in the \os\ panel shows the predicted contribution from a conductive
  interface with a velocity and width defined by the \cf\ and \sif\
  profiles, and with a typical \cf/\os\ ratio.}
\end{figure}

\begin{figure}[!ht]
\epsscale{0.85}
\plotone{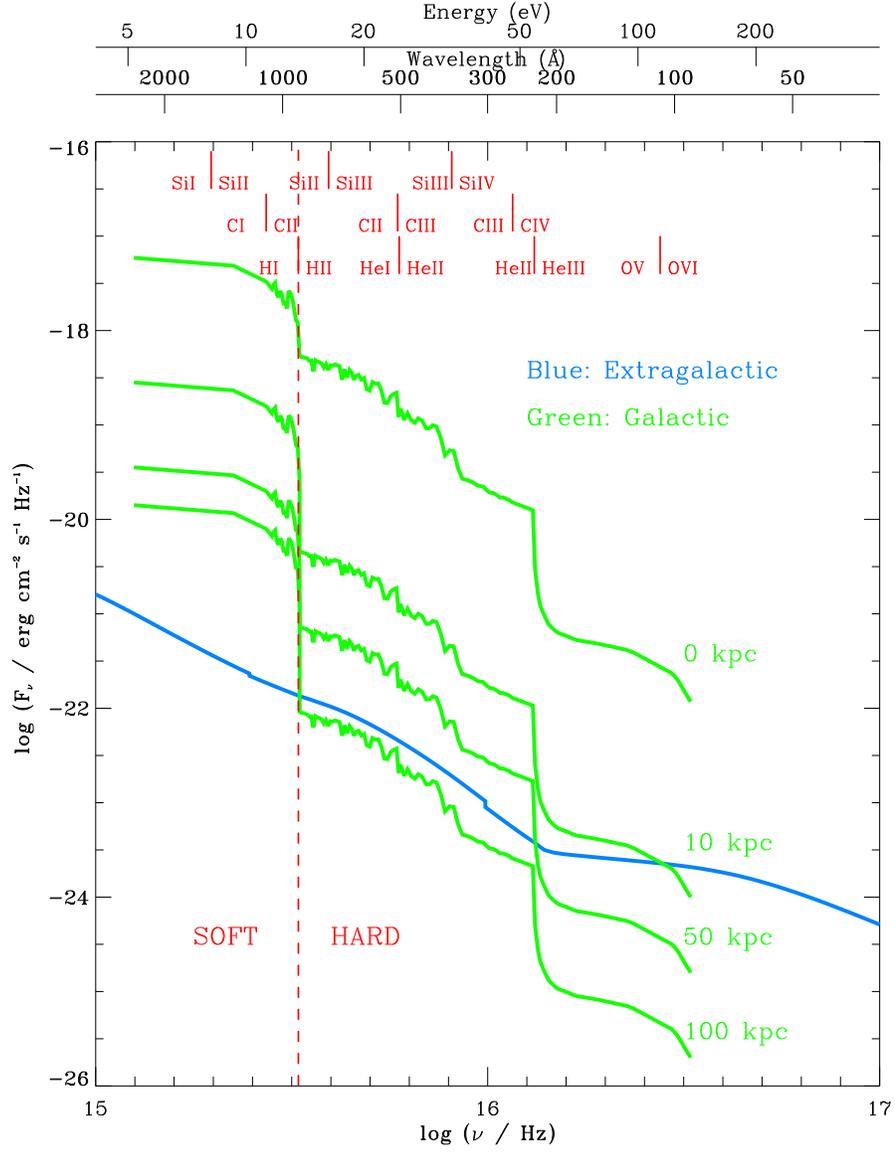}
\caption{Radiation fields incident upon the HVCs in our CLOUDY
  calculations. The blue line shows the estimated
  extragalactic background flux, where $F_\nu=4\pi J_\nu$ and
  $J_\nu$ is taken from \citet{HM96}. The green line shows our new estimate of
  the flux of escaping Galactic radiation, at distances of 0, 10, 50, and
  100\,kpc along the \pg\ sight line ($b=+52\degr$). 
  %The halo opacity is
  %higher for hard photons than for soft photons, so the scaling factors
  %(determined using Figure 9) are different for each band. 
  Note the substantial drop at 
  54\,eV in the Galactic spectrum, caused by the \ion{He}{2} edge in
  hot stars. Other ionization edges are marked at the top of the plot.}   
\end{figure}

\begin{figure}[!ht]
\epsscale{0.85}
\plotone{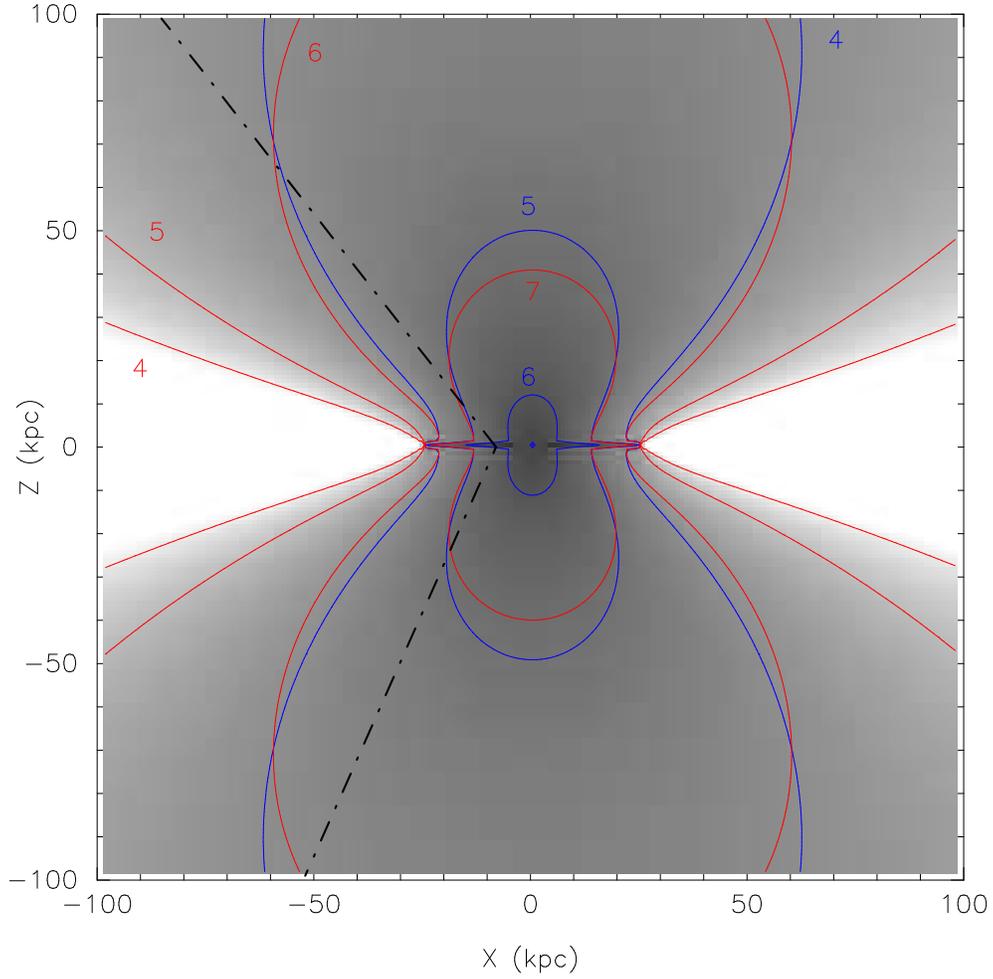}
\caption{Contour plot showing the flux of Galactic radiation
  as a function of location relative to the Galactic disk in our new
  disk-halo model.%, incorporating both geometric and opacity effects. 
  The blue contours show the lines of constant hydrogen ionizing
  flux ($\Phi_{<912}=\int_0^{912}(F_{\lambda}/h\nu){\mathrm d}\lambda$),
  and the red contours shows the lines of constant
  non-hydrogen ionizing
  flux ($\Phi_{>912}=\int_{912}^{2460}(F_{\lambda}/h\nu){\mathrm d}\lambda$). 
  The numbers next to the contours are the logarithm
  of $\Phi_{<912}$ in units of \sqcm\,s$^{-1}$.
  The Galactic ionizing field dominates the extragalactic ionizing
  background inside the line defined by 
  log\,$\Phi=4$. The dot-dashed lines show the sight lines of interest
  in this paper (\pg\ above the plane and \he\ below it). This map is
  used to scale the normalization of the 
  Galactic disk ionizing spectrum shown in Figure 8 for use in our
  photoionization modeling.}
\end{figure}

%\begin{figure}[!ht]
%\epsscale{1.15}
%\plottwo{f10a.eps}{f10b.eps}
%\caption{Ionic column densities as a function of ionization parameter
%  expected if HVCs are 
%  subjected to the EGB radiation field (left), or MW field at
%  10\,kpc (right), for a typical HVC with log\,$N$(\hi)=16.30 and
%  [$Z$/H]=$-$0.50. A given ionization parameter maps onto a different
%  total hydrogen density in each case, as shown on the upper
%  x-axes. Very low amounts of \os\ and \cf\ are produced by either 
%  radiation field unless the ionization parameter is very high.}
%\end{figure}

\begin{figure}[!ht]
\epsscale{0.8}
\plotone{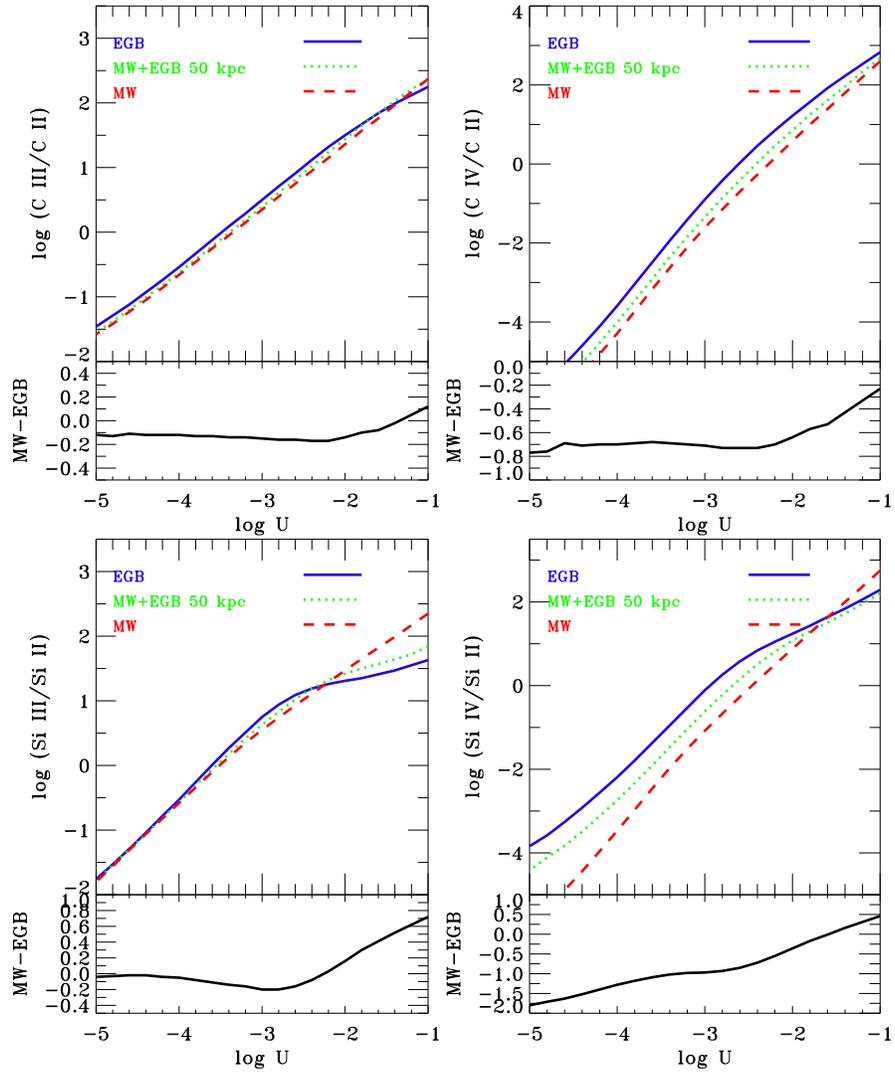}
\caption{The behavior of various column density ratios of carbon and
  silicon as a function of ionization 
  parameter, for HVCs immersed in three radiation fields: the EGB, the MW
  (any distance), and MW+EGB at 50\,kpc toward
  \he. The logarithmic differences in the ionic ratios between the
  pure MW and pure EGB fields are shown in the panels labelled MW--EGB. 
  %At low ionization parameter (log\,$U<-2.0$), the MW field produces  
  %relatively less \cf\ and \sif\ than the EGB field. At high
  %ionization parameter (log\,$U>-2.0$), the MW field produces 
  %relatively more \ct, \sit, \cf, and \sif\ than the EGB field. 
  Under the assumption of pure photoionization, these curves
  can be used to convert an observed ionic ratio to an implied
  ionization parameter, for a given radiation field.}
\end{figure}

\begin{figure}[!ht]
\epsscale{1.15}
\plottwo{f11a.eps}{f11b.eps} 
\caption{Example CLOUDY run showing the fits to observed column densities in
  HVC2 toward \pg, for two radiation fields: EGB (left) and MW+EGB at
  10\,kpc (right). 
  %log\,$N$(ion) is plotted against log\,$U$ and the inferred value of
  %log\,$n_{\mathrm{H}}$. 
  For both radiation fields
  a good fit can be found to the \cw, \ct, \siw, and \sit\ data, but not
  the \cf, \sif, or \os. The best-fit log\,$U$ and [$Z$/H] and their
  errors are derived using a
  chi-squared minimization process,
  %to the singly and doubly ionized species, 
  shown in the grayscale on the bottom panel. The contours show the
  regions of 70, 95, and 99\% confidence.
  %$U$ is then
  %used to derive the
  %article density, pressure, and cloud size annotated near the top of the
  %plots. In the Milky Way field, the cloud has a higher
  %inferred density and pressure, and smaller physical size.
   Results from models of this kind to all HVC components toward \he\ and \pg\
  are given in Table 7.} 
\end{figure}

\begin{figure}[!ht]
\epsscale{1.0}
\plottwo{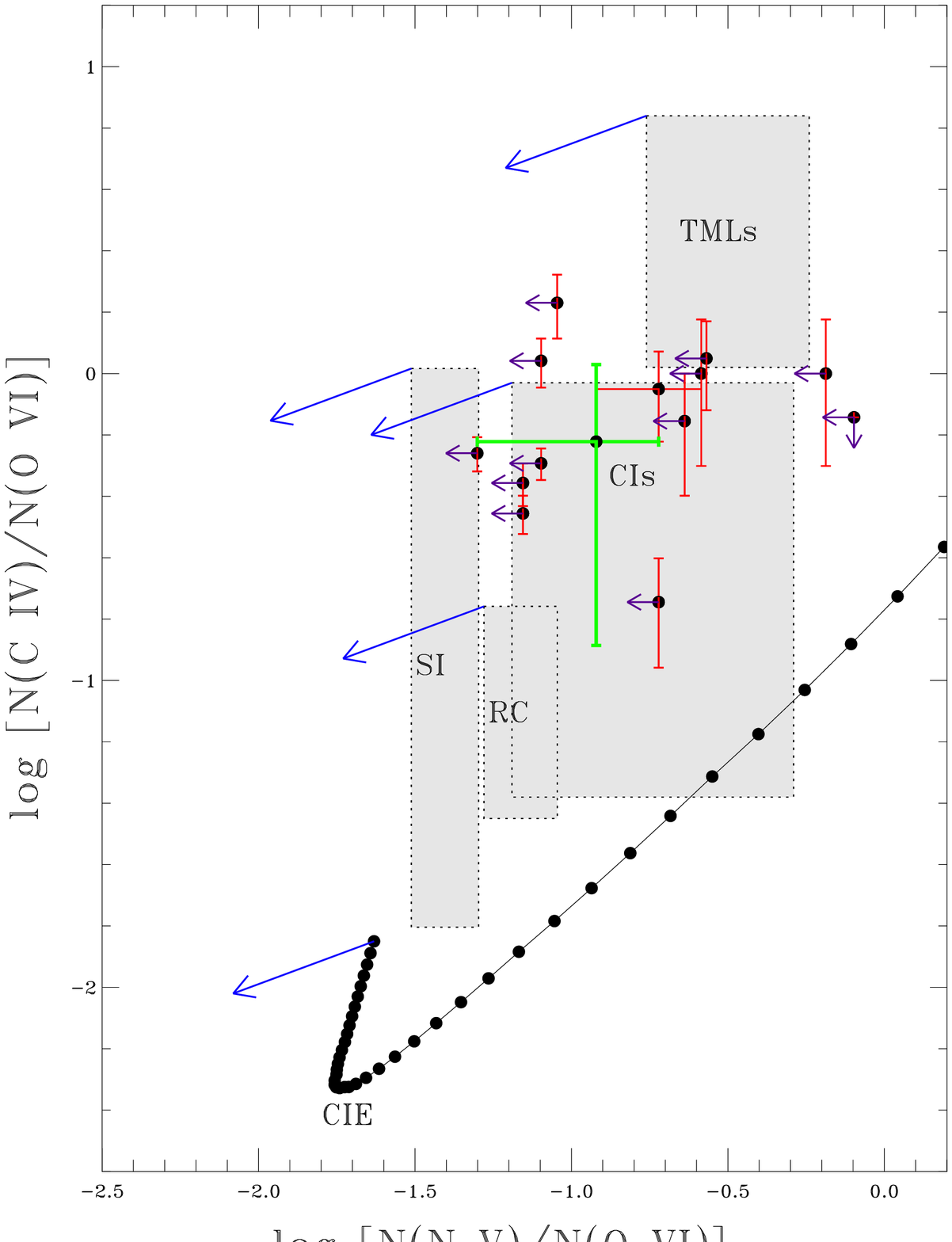}{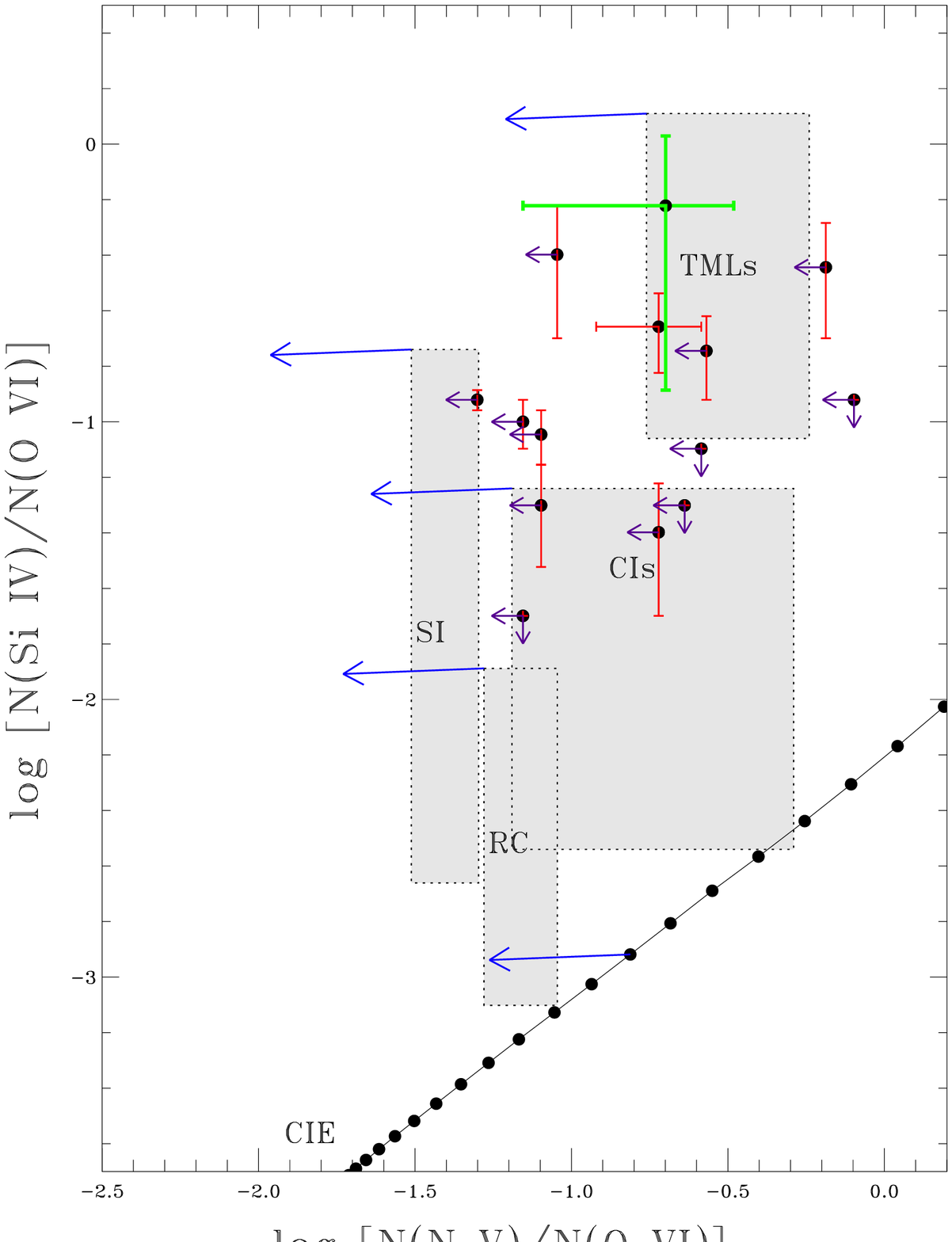}
\caption{High-ion column density ratios measured in HVCs (red points,
  purple upper limits),
  the Galactic halo average \citep[green point;][]{Zs03}, and the
  predictions of various models (gray boxes). CIs = conductive
  interfaces; CIE =  collisional ionization equilibrium; TMLs =
  turbulent mixing layers; 
  RC = radiative cooling; SI = shock ionization. The blue arrows show
  the effect of correcting the model predictions with solar abundance
  ratios onto a
  Magellanic Stream set of abundances. See \citet{Fo04} for a detailed
  explanation of the model parameters.}
\end{figure}

\end{document}